# Realistic micromagnetic description of all-optical ultrafast switching processes in ferrimagnetic alloys

V. Raposo[1,*], F. García-Sánchez[1], U. Atxitia[2], and E. Martínez[1,+]

1. *Applied Physics Department, University of Salamanca.*
2. *Dahlem Center for Complex Quantum Systems and Fachbereich Physik.*

**\*,+**: Corresponding authors: victor@usal.es, edumartinez@usal.es

**Abstract**

Both helicity-independent and helicity-dependent all-optical switching processes driven by single ultrashort laser pulse have been experimentally demonstrated in ferrimagnetic alloys as GdFeCo. Although the switching has been previously reproduced by atomistic simulations, the lack of a robust micromagnetic framework for ferrimagnets limits the predictions to small nano-systems, whereas the experiments are usually performed with lasers and samples of tens of micrometers. Here we develop a micromagnetic model based on the extended Landau-Lifshitz-Bloch equation, which is firstly validated by directly reproducing atomistic results for small samples and uniform laser heating. After that, the model is used to study ultrafast single shot all-optical switching in ferrimagnetic alloys under realistic conditions. We find that the helicity-independent switching under a linearly polarized laser pulse is a pure thermal phenomenon, in which the size of inverted area directly correlates with the maximum electron temperature in the sample. On the other hand, the analysis of the helicity-dependent processes under circular polarized pulses in ferrimagnetic alloys with different composition indicates qualitative differences between the results predicted by the magnetic circular dichroism and the ones from inverse Faraday effect. Based on these predictions, we propose experiments that would allow to resolve the controversy over the physical phenomenon that underlies these helicity-dependent all optical processes.

# I. INTRODUCTION

All optical switching (AOS) refers to the manipulation of the magnetic state of a sample through the application of short laser pulses. The discovery of subpicosecond demagnetization of a nickel sample [1] upon application of a short laser pulse, ranging from tens of femtosecond to several picoseconds, opened the path for other experiments to manipulate the magnetization using ultrashort laser pulses in ferromagnetic [2,3], synthetic antiferromagnetic [4–6] and ferrimagnetic materials [7–9]. While the AOS in ferromagnetic materials is usually described by the Magnetic Circular Dichroism (MCD) [10,11] or the Inverse Faraday Effect (IFE) [12–14], and it requires multiple shots of circularly polarized laser pulses [15], the inversion of the magnetization of ferrimagnetic materials can be achieved by single-shot pulse [16], even with linear polarization. In these Helicity-Independent AOS (HI-AOS) processes, the reversal takes place as the two antiferromagnetically coupled sublattices demagnetize at different rates when submitted to a laser pulse of adequate duration and energy. Since exchange processes conserve total angular momentum, the system transits through a ferromagnetic-like state despite being ferrimagnetic at the ground state [17]. The switching of the magnetization is completed when the sublattices relax back to their thermodynamic equilibrium [18]. On the other hand, several experimental studies [7,9] have also observed that the magnetic state of ferrimagnetic alloys can be also reversed under circular polarized laser pulses within a narrow range of laser energies, resulting in a Helicity-Dependent AOS (HD-AOS) which could be useful to develop ultrafast magnetic recording devices purely controlled by optical means. While the single-shot HI-AOS can be caused by the strong non-equilibrium due to the heating induced by the laser pulse, the physical mechanisms behind the HD-AOS are still not completely understood, and several works participated by the same authors ascribe it either to the IFE [9] or the MCD [8]. Although several attempts have been performed to explain such AOS processes, a realistic numerical description of experimental observations is still missing. Indeed, some theoretical studies usually adopt an atomistic description which is limited to small samples [16,19], with dimensions at the nanoscale, well below the size of the experimentally studied samples, with lateral sizes of several hundreds of microns. Such atomistic approach cannot describe the non-uniform heating caused by laser beams of several microns, so it does not predict some multidomain patterns typically observed in the experiments [9]. On the other hand, other numerical attempts have been carried out by describing the ferrimagnetic alloy as an effective ferromagnetic sample, without considering the individual nature of the two sublattices forming the ferrimagnet [9]. Although these micromagnetic studies predict some features of the AOS processes, so far, the structure of the ferrimagnetic alloys has not been taken into account to investigate the reversal of magnetic samples of micrometer size under realistic excitation conditions. As the switching happens due to angular momentum transfer between sublattices, something impossible to account for within an effective ferromagnetic description, it is needed to develop studies considering the two sublattice nature of the such alloys to naturally evaluate their role on the reversal processes.

Here we present a micromagnetic framework that is able to reproduce accurately the atomistic results of the laser-induced switching by the extension of the conventional Landau-Lifshitz-Bloch (LLB) model for ferrimagnets [20,21]. Note that the conventional LLB model does not allow to accurately describe AOS as indicated in [21], and here we extend it to solve this limitation. However, and differently from atomistic simulations, which are limited to small samples at the nanoscale submitted to uniform laser heating, our micromagnetic formalism allows us to realistically describe AOS experimental observations by directly evaluating extended samples at the microscale and non-uniform energy absorption from the laser pulse. The procedures here developed are essential to understand the physical aspects underlying these experiments, and will be useful for the future development of novel ultra-fast devices based on these AOS processes. After presenting and validating both the atomistic and the extended micromagnetic models for sample size of tens of nanometers, the upper size limit of the atomistic spin models, we describe the results for HI-AOS processes in realistic samples at the microscale for a typically ferrimagnetic alloy (GdFeCo). Later on, we focus our attention to the description of the HD-AOS processes by exploring the role of the IFE and MCD separately for two different ferrimagnetic alloys where the relative composition is slightly varied. Our results allow us to suggest future experiments which could be useful to infer the dominance of the IFE or the MCD in single-shot HD-AOS in ferrimagnetic alloys.

## II. ATOMISTIC AND MICROMAGNETIC MODELS

Typical ferrimagnetic (FiM) samples formed by a Transition Metal (TM:Co, CoFe) and a Rare Earth (RE:Gd) are considered here. Square samples in the $xy$ plane with side length $\ell$ and with thickness $t_{FiM} = 5.6$ nm are studied. At atomistic level the FiM sample is formed by a set of coupled spins, and the magnetization dynamics is described the Langevin-Landau-Lifshitz-Gilbert equation

$$\frac{\partial \vec{S}_i}{\partial t} = -\frac{\gamma_0}{(1+\lambda^2)}\{\vec{S}_i \times (\vec{H}_i + \vec{H}_{th,i}) + \lambda \vec{S}_i \times [\vec{S}_i \times (\vec{H}_i + \vec{H}_{th,i})]\} \quad (1)$$

where $\vec{S}_i$ is the localized magnetic moment and $\vec{H}_i$ is the local effective field including intra- and inter-lattice exchange and anisotropy contributions. $\vec{H}_{th,i}$ is the local stochastic thermal field. $\gamma_0$ and $\lambda$ are the gyromagnetic ratio and the damping parameter respectively [22]. Except the contrary is indicated, typical parameters of $Gd_x(CoFe)_{1-x}$ with relative composition $x = 0.25$ were considered [22]. See Supplemental Material Note SN1 [23] for further details, including material and numerical parameters.

Starting from an initial uniform state of the FiM with the spins of the two sublattices antiparallelly aligned each other along the easy axis $z$, a laser pulse is applied, and the irradiated sample absorbs energy from the laser pulse. The laser spot is assumed to have a spatial Gaussian profile ($\eta(r)$), with $r_0$ being the radius spot ($d_0 = 2r_0$ is the full width at half maximun, FWHM). Its temporal profile ($\xi(t)$) is also Gaussian, with $\tau_L$ representing the pulse duration (FWHM). The absorbed power density can be expressed as $P(r,t) = Q\eta(r)\xi(t)$ where $\eta(r) = \exp[-4\ln(2)\, r^2/(2r_0)^2]$ is the spatial profile with $r = \sqrt{x^2 + y^2}$ being the distance from the center of the laser spot, and $\xi(t) = \exp[-4\ln(2)\,(t-t_0)^2/\tau_L^2]$ the temporal profile. $Q$ is the maximum value of the absorbed power density reached at $t = t_0$ just below the center of the laser spot.

Laser pulse heats the FiM sample, and consequently, it is transiently dragged into a non-equilibrium thermodynamic state, where its magnetization changes according to the temperature dynamics. The temperature evolution is described by the Two Temperatures Model (TTM) [9,24] in terms of two subsystems: the electron ($T_e = T_e(\vec{r},t)$) and the lattice ($T_l = T_l(\vec{r},t)$),

$$C_e \frac{\partial T_e}{\partial t} = -k_e \nabla^2 T_e - g_{el}(T_e - T_l) + P(r,t) - C_e \frac{(T_e - T_R)}{\tau_D} \quad (2)$$

$$C_l \frac{\partial T_l}{\partial t} = -g_{el}(T_l - T_e) \quad (3)$$

where $C_e$ and $C_l$ denote the specific heat of electrons and lattice subsystems, respectively. $k_e$ is the electronic thermal conductivity. $g_{el}$ is a coupling parameter between the electron and lattice subsystems, and $\tau_D$ is the characteristic heat diffusion time to the substrate and the surrounding media [25]. Conventional values were adopted (see [1,16,26] and Supplemental Material Note SN1 [23]).

The approach that consists on solving Eq. (1) coupled to Eqs. (2)-(3) is named as **Atomistic Spin Dynamics** (ASD), and due to computational restrictions, its numerical solution is limited to small samples at the nanoscale ($\ell \lesssim 100$ nm, see Supplemental Material Note SN2 [23]). While ASD predicts the single-shot switching in small FiM nano-samples [16,22,27], the lack of a realistic micromagnetic framework for micro-size samples and non-unifom laser spot limits the description of many experimental works [9]. In particular, the appearance of central regions with a multi-domain demagnetized patterns [28], or the observation of rings of switched magnetization under irradiation with lasers of tens of micrometers [29] cannot be reproduced by ASD due to such computing limitations. In order to overcome the ASD limitations, here we develop an extended continuous micromagnetic model that describes the temporal evolution of the reduced local magnetization $\vec{m}_i(\vec{r},t)$ of each sublattice $i$:RE,TM based on the conventional ferrimagnetic Landau-Lifshitz-Bloch (LLB) Eq [21,30],

$$\frac{\partial \vec{m}_i}{\partial t} = -\gamma'_{0i}(\vec{m}_i \times \vec{H}_i) + $$
$$-\frac{\gamma'_{0i}\alpha_i^\perp}{m_i^2}\vec{m}_i \times [\vec{m}_i \times (\vec{H}_i + \vec{\xi}_i^\perp)] + \quad\quad (4)$$
$$+\frac{\gamma'_{0i}\alpha_i^\parallel}{m_i^2}(\vec{m}_i \cdot \vec{H}_i)\vec{m}_i + \vec{\xi}_i^\parallel$$

where $\vec{H}_i = \vec{H}_i(\vec{r},t)$ is the local effective field on sublattice magnetic moment $i$ at location $\vec{r}$ of the FiM sample, $\alpha_i^\parallel$ and $\alpha_i^\perp$ are the longitudinal and perpendicular damping parameters, and $\vec{\xi}_i^\parallel$ and $\vec{\xi}_i^\perp$ are the longitudinal and perpendicular stochastic thermal fields. Details of the LLB model can be found in [21,30] and in Supplemental Material Note SN2 [23]. Contrary to the ASD, where the spatial discretization is imposed by the atomistic scale ($a = 0.35$ nm), within the micromagnetic model the sample is discretized in elementary cells with dimensions of $\Delta x = \Delta y \sim 1$nm and $\Delta z = t_{FiM}$. Therefore, it is possible to numerically evaluate, with manageable computing effort, extended samples at the microscale ($\ell \sim 100$ μm), three orders of magnitude larger than the ones which can be dealt with the ASD model.

Numerically solving Eq. (4) coupled to Eqs. (2)-(3) under ultra-short laser pulses provides a micromagnetic description of several AOS processes in ferromagnetic systems [13]. However, when dealing with ferrimagnetic samples we checked that some disagreement with the predictions of the ASD model were observed (see Supplemental Material Note SN3 [23]), which are related to the lack of a proper description of the angular moment exchange between sublattices during the non-equilibrium transient state promoted by the laser pulse. Indeed, magnetization dynamics in FiMs is driven by dissipative processes of relativistic and exchange nature. The relativistic ones allow exchange of angular momentum between the magnetization and the lattice degree of freedom due to the spin-orbit coupling between them, and are phenomenologically described by the usual damping terms in the LLB Eq. (4). Additionally, in multisublattice magnets as FiMs, another different pathway opens local exchange of angular momentum between both sublattices of the FiM, and to account for it, the LLB Eq. (4) has to be enhanced by an additional exchange relaxation torque [18,31–34]. The simplest model to describe the sublattice-specific magnetization dynamics in FiMs, was derived from Onsager's relations [31] within a macrospin approach based on a microscopic spin model. In this simplified description, the magnetization dynamics of sublattice $i$ can be expressed as $\frac{1}{\gamma_{0i}}\frac{dm_i}{dt} = \alpha_i H_i + \alpha_{ex}\left(\frac{\mu_i}{\mu_j}H_i - H_j\right)$, where $i,j$: RE, TM, $\mu_i$ and $H_i$ are the magnitude of the magnetic moment and the effective field acting on macrospin of sublattice $i$ respectively. The relativistic relaxation parameter in this model, $\alpha_i$, corresponds to the longitudinal damping parameter in the LLB Eq. and it depends on the temperature of the thermal bath to which angular momentum and energy is dissipated. Differently, it is assumed that the exchange relaxation parameter $\alpha_{ex}$, only depends on the non-equilibrium sublattice magnetizations, $\alpha_{ex} = \alpha_{ex}(m_i, m_j)$. Considering that exchange relation rate should be symmetric with respect to the sublattice index, $\alpha_{ex}(m_i, m_j) = \alpha_{ex}(m_j, m_i)$, a simple functional fulfilling these heuristic conditions yields to $\alpha_{ex}(m_i, m_j) = \lambda_{ex}\frac{m_i + (x_j\mu_j/x_i\mu_i)m_j}{m_i m_j}$ where $\lambda_{ex}$ is a phenomenological parameter representing the exchange relaxation rate and $x_i$ the concentration of each specimen $i$. Inspired by this two sublattice phenomenological model based on Onsager's relations, here we add an additional torque $\vec{\tau}_i^{NE}$ to the micromagnetic LLB Eq. (4) that accounts for non-equilibrium magnetic moment exchange between sublattices, and becomes crucial to describe AOS ultra-fast switching in FiMs under realistic conditions. The torque reads as

$$\vec{\tau}_i^{NE} = \gamma'_{0i} \lambda_{ex} \alpha_i^{\|} \frac{x_i \mu_i m_i + x_j \mu_j m_j}{\mu_i m_i \mu_j m_j} \left( \mu_i \vec{H}_i^{\|} - \mu_j \vec{H}_j^{\|} \right) \quad (5)$$

where and $\vec{H}_i^{\|}$ and $\vec{H}_j^{\|}$ are the longitudinal effective fields for each lattice $i$:RE,TM [21], $x_i \equiv x$ and $x_j = 1 - x_i = 1 - x$ are the concentrations of each specimen, and $\lambda_{ex}$ is a parameter representing the exchange relaxation rate [18]. By including Eq. (5) in the RHS of Eq. (4), and numerically solving it coupled to TTM Eqs. (2)-(3), we can provide a realistic description of the magnetization dynamics in FiM systems under ultra-short laser pulses. In what follows, we refer to this formalism as the ***extended micromagnetic LLB model*** (eLLB), to distinguish it from the ***conventional LLB model*** (LLB) when $\vec{\tau}_i^{NE} = 0$. See Supplemental Material Note SN2 [23] for the rest of details.

**III. RESULTS AND DISCUSSION**

Before presenting the predictions of the extended micromagnetic model for realistic FiM samples and laser beams at the microscale, here we firstly compare the results obtained from the extended LLB model (eLLB) to the ones resulting from the atomistic spin dynamics simulations (ASD) for a small FiM dot at the nano-scale ($\ell \approx 25$ nm). As typical laser spots have radius of $r_0 \sim 1$ μm $- 10$ μm or even larger, we assume here that the power absorbed by the FiM dot the from the laser pulse is uniform, that is, $\eta(r) = 1$. The pulse duration is $\tau_L = 50$ fs. Typical results showing the temporal evolution of the out-of-plane averaged magnetization ($m_z^i$) for each sublattice ($i$:TM (red), RE (blue)) are shown in FIG. 1 for two different values of $Q$.

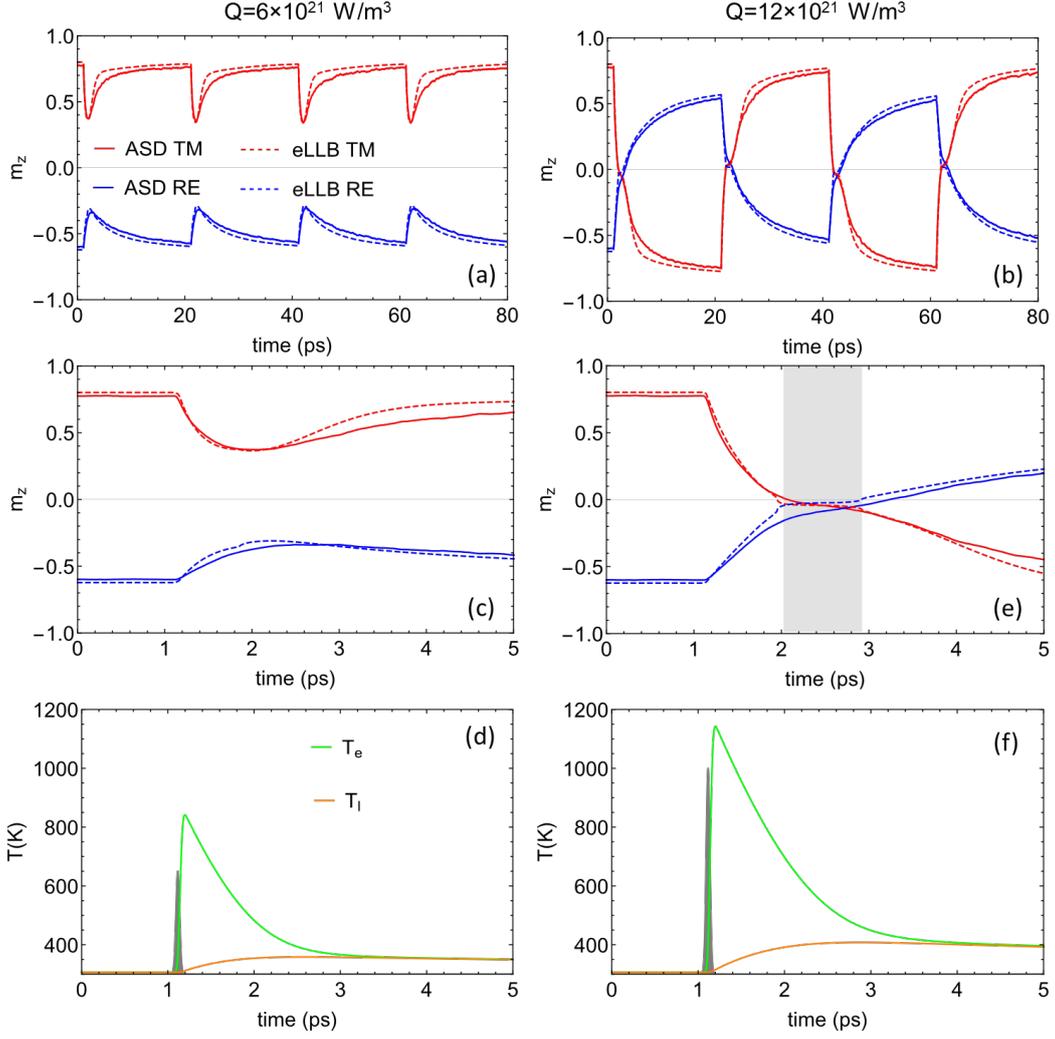

FIG. 1. Comparison between atomistic simulation (ASD, solid lines) and micromagnetic model (eLLB, dashed lines) results for the temporal evolution of the out-of-plane magnetization for the transition metal (TM, red) and the rare earth (RE, blue) under two different laser power densities: (a) $Q = 6 \times 10^{21}$ W/m$^3$ and (b) $Q = 12 \times 10^{21}$ W/m$^3$. Four consecutive laser pulses with $\tau_L = 50$ fs are applied every 20 ps. (c) shows a detailed view of first pulse switching event as in (a), while (d) shows temporal evolution of the electron ($T_e$) and lattice ($T_l$) temperatures for $Q = 6 \times 10^{21}$ W/m$^3$. (e) and (f) corresponds to $Q = 12 \times 10^{21}$ W/m$^3$. Shaded interval in (e) shows the transient ferromagnetic state. The pulse length is $\tau_L = 50$ fs. The eLLB results were obtained with $\lambda_{ex} = 0.013$.

A remarkable agreement between both ASD and eLLB models with similar dynamics for both sublattices is observed in FIG. 1(a) and (b). For low $Q$ values (FIG. 1(a) and (c)) there is no switching, but when $Q$ increases above a threshold, which depends on the pulse length ($\tau_L$), the deterministic AOS is predicted by both ASD and eLLB models (FIG.1(b) and (e)). It is important to note that similar switching was also obtained within the deterministic eLLB framework, that is, in the absence of thermal fluctuations ($\vec{\xi}_i^\perp = \vec{\xi}_i^\parallel = 0$ in Eq. (4), see FIG. S4(b) in Supplemental Material Note SN3 [23]). On the contrary, the conventional LLB model (LLB, $\vec{\tau}_i^{NE} = 0$) fails to reproduce the switching of FIG. 1(b) (see FIG. S4(c)-(d) in Supplemental Note SN3 [23]). FIG. 1(c) shows the details of the temporal evolution $m_z^i$ for $Q = 6 \times 10^{21}$ W/m³ during the first laser pulse, while the corresponding evolutions of $T_e$ and $T_l$ are depicted in FIG. 1(d), which also shows the laser pulse. Corresponding results for $Q = 12 \times 10^{21}$ W/m³ are shown in FIG. 1(e) and (f) respectively. For $Q = 6 \times 10^{21}$ W/m³, the electron temperature reaches a peak maximum value of $T_e \approx 850$ K at the end of the laser pulse, but this is not enough to achieve the switching. For $Q = 12 \times 10^{21}$ W/m³, $T_e$ reaches a peak of $T_e \approx 1150$ K and switching takes place. Notice that this value is well above the Curie temperature ($T_C \approx 600$ K), and therefore the system needs to be significantly heated above the Curie threshold to achieve the deterministic AOS in FiM. These processes are explained by the different demagnetization rates of the RE and TM sublattices, that lead to a transient ferromagnetic aligment. Such transient ferromagnetic state is observed during a short transient (see shaded interval in FIG.1(e)), and it is only present when the system is far away from the thermodynamic equilibrium, as caused by the ultrafast laser heating. Except the contrary is indicated, all eLLB results were obtained with $\lambda_{ex} = 0.013$, (see FIG. S5 and its corresponding discussion in Supplemental Material Note SN3(c) [23] for results with other values of $\lambda_{ex}$).

***Helicity-Independent All Optical Switching* (HI-AOS)**. Once validated the eLLB formalism by reproducing the ASD results for small nano-samples under uniform linearly-polarized laser pulses, we can now use it to explore the influence of the laser duration ($\tau_L$) and maximum absorbed power density ($Q$) in realistic extended samples at the microscale ($\ell \sim 10$ μm). This is illustrated in the phase diagram of FIG. 2(a), which shows the final state under a single linearly-polarized laser pulse starting from a uniform state of the FiM. White color indicates the combinations of ($Q, \tau_L$) where the sample returns to the original state after the pulse (no-switching). The blue region corresponds to combinations of ($Q, \tau_L$) presenting deterministic HI-AOS after each pulse, and red corresponds to combinations of ($Q, \tau_L$) resulting in a final demagnetized multidomain configuration. It is noted that there is a correlation between the final state and the maximum electron temperature reached in the sample, which is shown by the overlapping solid black lines in FIG. 2(a). As it is clearly observed, solid lines coincide with boundaries between the three possible behaviors already discussed. Indeed, the transition between no-switching (white) to the deterministic switching range (red) is limited by the ~1000 K curve, whereas the transition to the thermal demagnetization (blue) occurs when $T_e \gtrsim 1400$ K, as shown in FIG. 2(a). Instead of $Q$, the information collected in the phase diagram of FIG. 2(a) could be also presented in terms of the laser fluence ($F \equiv Q \, \tau_L \, t_{FiM}$), as it is done in FIG. S6 of Supplemental Material Note SN4 [23]. Note, that such phase diagram is also in good qualitative agreement with recent experimental observations [35].

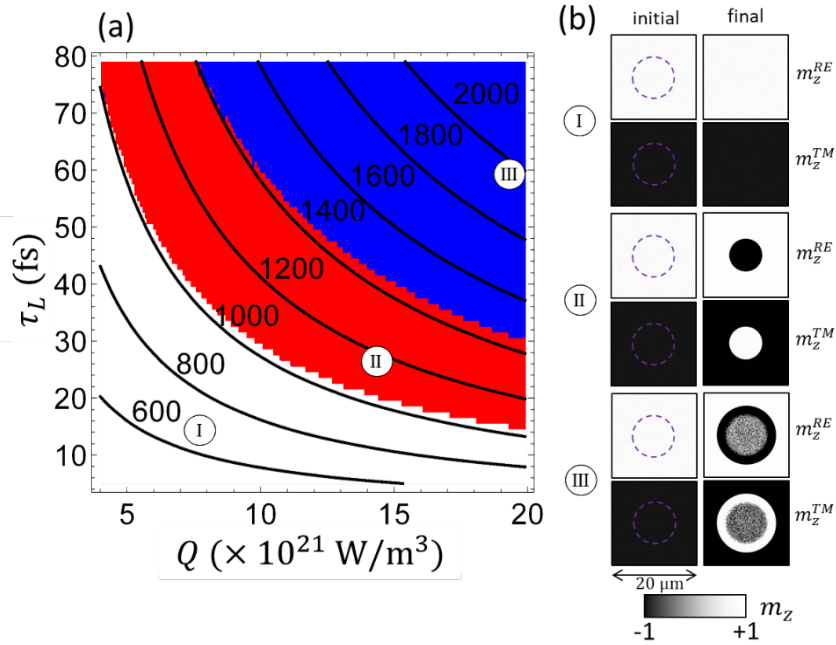

FIG. 2. (a) Phase diagram of the final state as a function of $Q$ and $\tau_L$ for a small nano-sample $\ell = 25$ nm under uniform laser heating ($\eta(r) = 1$). White, red and blue colors represent no-switching, deterministic switching and thermal demagnetization behaviors respectively. Solid lines are isothermal curves showing the maximum electron temperature ($T_e$) reached due to the pulse. (b) Typical micromagnetic snapshots of the initial and final magnetization of RE and TM for three combinations of $(Q, \tau_L)$. I: $(8 \times 10^{21}$ W/m$^3$, 20 fs), II: $(15 \times 10^{21}$ W/m$^3$, 30 fs) and III: $(20 \times 10^{21}$ W/m$^3$, 60 fs). Here extended samples ($\ell = 20$ μm) with a laser spot of $d_0 = \ell/2$ were considered. Dashed purple lines in the images of the initial state indicate the FWHM of the laser spot. The results of the phase diagram (a) coincide with (b) for magnetization at the center of the laser spot.

The main advantage of the extended eLLB model over ASD simulations is that it allows us to explore realistic samples and laser beams with dimensions that are not accessible with ASD models. eLLB model (Eqs. (4) and (2)-(3)) has been used to simulate samples with lateral size of $\ell = 20$ μm. From now on, the spatial Gaussian dependence of the laser beam is considered ($\eta(r) = \exp[-4\ln(2)\,r^2/(2r_0)^2]$), with a laser spot diameter of $d_0 = 2r_0 = \ell/2$. Typical initial and final states corresponding to three representative combinations of $(Q, \tau_L)$ are shown in FIG. 2(b). Our micromagnetic simulations point out again that the three types of behaviors observed experimentally (see for example Fig. 4(a) in [9] or [6,28]) are also achieved under these realistic conditions, with samples and laser spots at the microscale. Notice that now the final magnetic state depends on the local position because the power absorption from laser pulse does. The final states depict a radial symmetry around the center of the laser spot, which coincides with the center of the FiM sample at $(x_c, y_c) = (0,0)$.

In order to further describe such spatial dependence, FIG. 3 plots the final state of $m_z^{TM}$ as a function of $x$ along the central line of the FiM sample ($y = 0$) for the same three representative combinations of $(Q, \tau_L)$ as in FIG. 2(b). The maximum electron temperature $T_e = T_e(x)$ is also plotted in top graphs by blue curves. Bottom graphs in FIG. 3 show the final state over the sample plane $(x, y)$. These graphs clearly correlate the local final magnetic state ($m_z^{TM}(x, y)$) with the maximum electron temperature $T_e(x, y)$. In the no-switching regime (I), the electron temperature does not reach 1000 K at any point. For combinations $(Q, \tau_L)$ as II, $T_e(x) \gtrsim 1000$ K is only reached in the central region, whose dimensions fit the local part of the sample that switches its magnetization. Note that $T_e$ remains below $T_e(x) \lesssim 1400$ K. Finally, the demagnetized case (multi-domain pattern, III) occurs in the part of the sample where $T_e \gtrsim 1400$ K, but deterministic switching is still obtained in the ring region, where $1000$ K $\lesssim T_e(x) \lesssim 1400$ K. Micromagnetic images (bottom graphs in FIG. 3) are in good agreement with typical experimental HI-AOS observations [6,28].

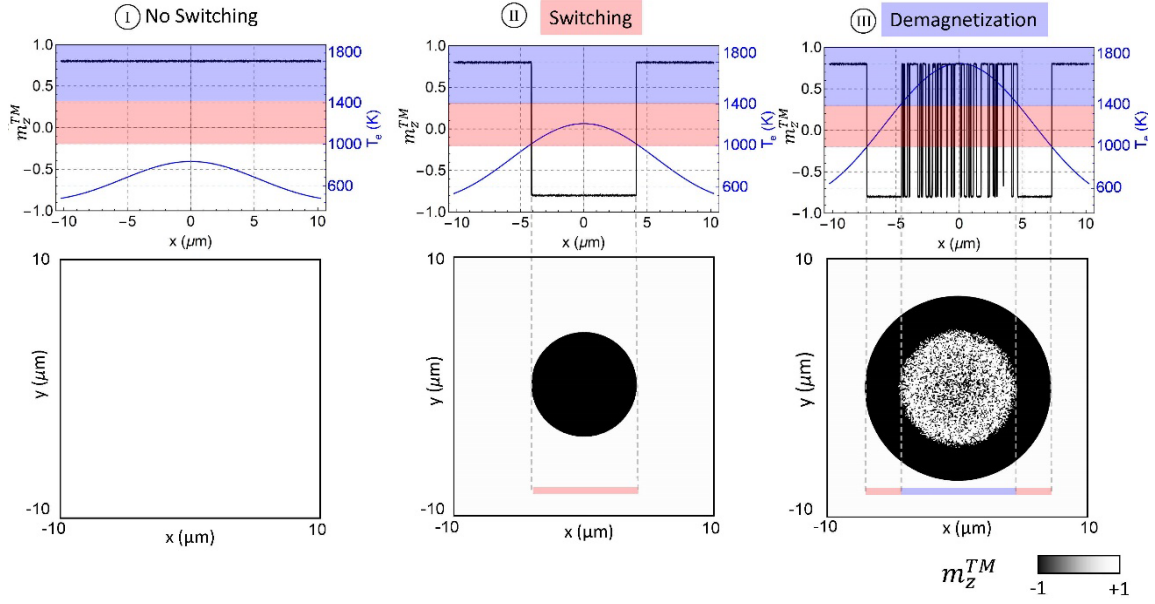

FIG. 3. Final out-of-plane magnetization ($m_z^{TM}(x)$) and maximum electron temperature ($T_e(x)$) as function of $x$ for $y = 0$ and for the three representative combinations of ($Q, \tau_L$) as in FIG. 2(b) (top graphs). The corresponding final states over the sample plane ($x, y$) are shown in bottom graphs.

The inferred correlation between the maximum electron temperature and the final magnetic state allows us to predict the size of the inverted region by studying the maximum electron temperature reached in the sample by using the TTM (Eqs. (2)-(3)) in combination with the switching diagram of FIG. 2(a). The radius of the switched area ($R_s$) calculated from micromagnetic simulations (dots, eLLB), and the one predicted by the TTM (lines, TTM) is shown in FIG. 4 as function of $\tau_L$ for two different values of $Q$ within the deterministic switching range (1000 K $\lesssim T_e(x) \lesssim$ 1400 K). Again, good agreement is obtained, a fact that points out that the origin of these HI-AOS processes under linearly polarized laser pulse is a purely thermal phenomenon. Indeed, as the local maximum electron temperature reached in the sample only depends on the absorbed power from the laser ($Q$) and the laser pulse length ($\tau_L$), the size of the inverted region can be directly obtained from the TTM by the condition of $T_e \gtrsim$ 1000 K.

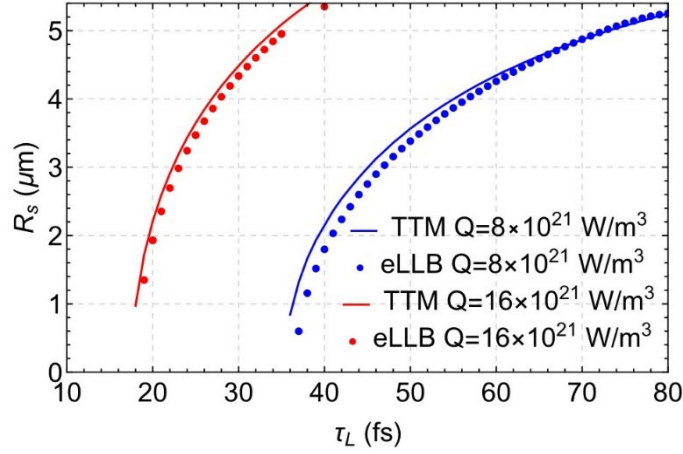

FIG. 4. Radius of the switched area ($R_s$) as a function of laser duration ($\tau_L$) for two values of the maximum absorbed power ($Q$). Dots are micromagnetic results from the extended eLLB model. Lines are predictions from the TTM, where $R_s$ is inferred from the condition that the local maximum electron temperature reaches $T_e \gtrsim 1000$ K.

*Helicity-Dependent All Optical Switching* **(HD-AOS)**. Previous results were carried out by applying laser pulses with linear polarization ($\sigma = 0$), and show that the HI-AOS can be achieved in a controlled manner with an adequate election of the laser power ($Q$) and duration ($\tau_L$): the magnetization switches its direction in the picoseconds range independently on the initial state. While this is interesting for toogle memory applications, the procedure to store and record a bit using linearly polarized laser pulses would still require two steps: ($i$) a preliminary reading operation of the magnetic state, and after that, ($ii$) deciding or not to apply the laser pulse depending on the preceeding state. This two-step procedure can be avoided by using circularly polarized laser pulses, resulting in Helicity-Dependent AOS processes (HD-AOS). However, as it was already commented the physics behind these HD-AOS observations still remains unclear, and both the Magnetic Circular Dichroism (MCD) [8,10] and the Inverse Faraday Effect (IFE) [12,14] have been suggested as responsible of the experimental observations. In what follows, we explore both mechanisms in a separated manner by including them in the extended micromagnetic model.

Let firstly consider the *Magnetic Circular Dichroism*. It has been suggested to be play a dominant a role on these HD-AOS processes in GdFeCo ferrimagnetic samples, which are known for its strong magneto-optical effect [8]. According to the MDC formalism, right-handed ($\sigma^+$) and left-handed ($\sigma^-$) circularly polarized laser pulses experience different refractive indices, and consequently a difference in energy absorption of the FiM sample for $\sigma^+$ and $\sigma^-$ pulses is expected. The MCD coefficient can be calculated from the total absorption for each polarization, resulting in MCD $\equiv k = (A_- - A_+)/(\frac{1}{2}(A_+ + A_-))$, where $A_\pm$ represent the total absorption for each polarization, ($\pm \equiv \sigma^\pm$). Indeed, the MCD makes the power absorbed by the sample ($P(r,t)$) to depend on the laser helicity ($\sigma^\pm = \pm 1$ for right-handed and left-handed circular helicities) and on the initial net magnetic state ($m_N(0) = M_S^{TM} m_z^{TM}(0) + M_S^{RE} m_z^{RE}(0)$), *up* (↑: $m_N(0) > 0$) or *down* (↓: $m_N(0) < 0$). Note that $m_z^{TM}(0) = \pm 1$ and $m_z^{RE}(0) = \mp 1$, whereas $M_S^{TM}$ and $M_S^{RE}$ are both positive. Under a right-handed laser pulse ($\sigma^+$), an initially *up* (*down*) magnetic state is expected to absorb more (less) energy than the initially *down* (*up*) state. Therefore, $P(r,t)$ in Eq. (2) is replaced by $\psi(\sigma^\pm, m_N) P(r,t)$ with $\psi(\sigma^\pm, m_N) = \left(1 + \frac{1}{2} k \sigma^\pm \text{sign}(m_N)\right)$ describing the different absorption power for *up* and *down* magnetization states as depending on the laser helicity. See further details on the implementation of the MCD in Supplemental Material Note SN5 [23].

We have evaluated the role of the MCD in the eLLB model with several values of the MCD coefficient ($k$). The isothermal curve delimiting the border between the no-switching and switching regimes now depends on the combination of helicity and initial net magnetic state (see such isothermal threshold curves for different values of the MCD coefficient in of SN5). Considering a realistic value of MCD $\equiv k \sim 2\%$, as estimated in [8], the electron temperature variation is quite small, typically a few units of K, and therefore small variations in the phase diagram are obtained with respect to the one for linearly-polarized pulses FIG. 2(a) (see also FIG. S7 in Supplemental Material Note SN5 [23]). However, when exciting with circular polarized pulses close to the no-switching/switching boundary, the FiM switches or not depending on the helicity and initial net state, only within a narrow interval of $Q$. This is represented in FIG. 5(a), where the HD-AOS is shown for $\tau_L = 50$ fs pulses with different $Q$ in a sample with $\ell = 20$ μm. Note that these results correspond to a FiM alloy Gd$_x$(FeCo)$_{1-x}$ with $x$=0.25, and that for this relative composition the RE is the dominant sublattice at room temperature, $M_S^{RE} > M_S^{TM}$ at $T = 300$ K (see inset of FIG. S3 of Supplemental Material [23] or FIG. 6(a)). No switching is achieved for low energy values (see left colum in FIG. 5(a) for $Q = 5.7 \times 10^{21}$ W/m$^3$). However, if $Q$ increases to $Q = 5.8 \times 10^{21}$ W/m$^3$, the system shows the so-called HD-AOS: if $m_z^{RE}$ is initially *down* (*up*), the reversal is only achieved for left-handed helicity, $\sigma^- = -1$ (right-handed helicity, $\sigma^+ = +1$). Consequently, the final state can be selected by chosing the laser helicity, which is relevant for ultrafast memory applications. It is important to note that this Helicity Dependent AOS is only obtained in a very narrow range of $Q$ around the Helicity Independent AOS boundary. Indeed, a small increase of the absorbed power results again in HI-AOS as the linear polarized case (see 3$^{rd}$ and 4$^{th}$ columns in FIG. 5(a) for $Q = 5.9 \times 10^{21}$ W/m³ and $Q = 9.0 \times 10^{21}$ W/m³). For high values of $Q$, the final state depicts a ring around a central demagnetized state, similar to HI-AOS case (see right column in FIG. 5(a) for $Q = 18 \times 10^{21}$ W/m³). In this case, the maximum electron temperature overcomes the $T_e \simeq 1400$ K threshold in the central region below the laser spot, resulting in a central demagnetized or multidomain state. However, the maximum $T_e$ remains within the range of HD-AOS deterministic switching (1000 K $\lesssim T_e(x) \lesssim$ 1400 K) in the ring around the central part. These micromagnetic predictions, including the narrow range of HD-AOS, are in good agreement with several experimental observations (see, for instance, Fig. 4 and 7(a) in [9], Fig. 1(b) in [4], or Fig. 3 in [36]).

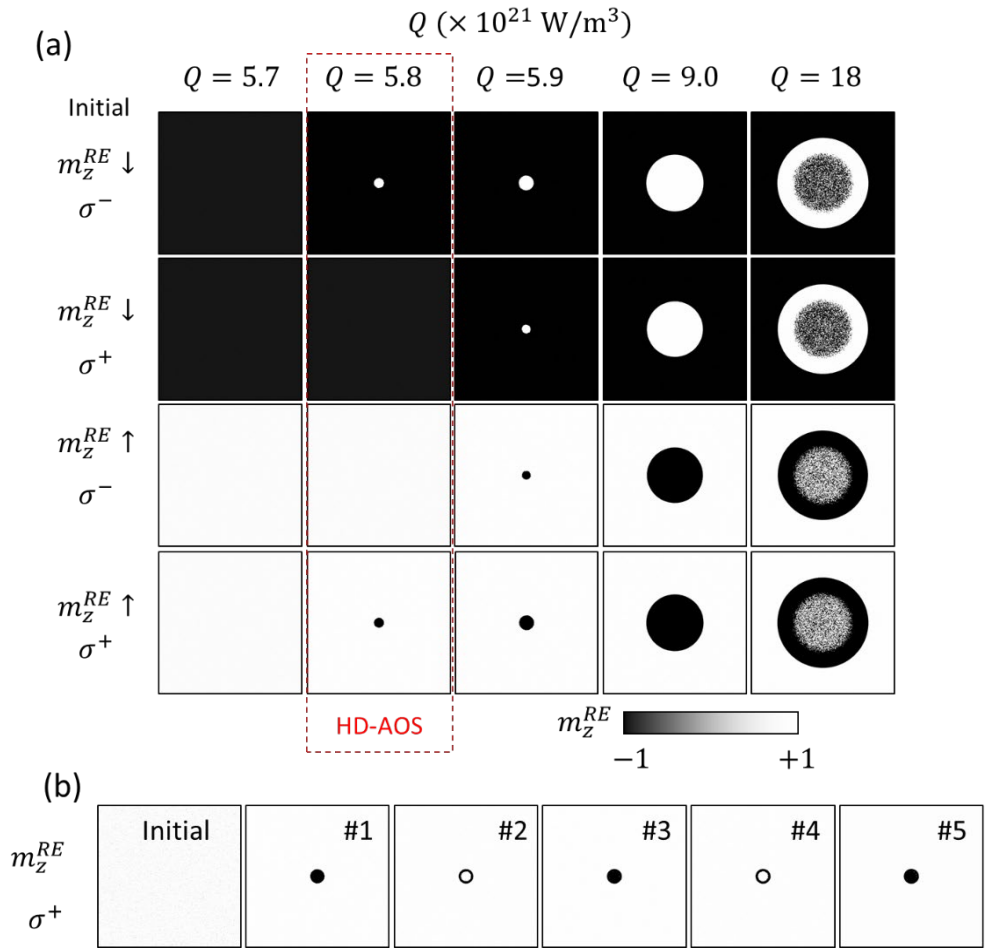

FIG. 5. Helicity-Dependent AOS predicted by the MCD. (a) Snapshots of the final RE magnetic state ($m_z^{RE}$) after a laser pulse of $\tau_L = 50$ fs for five different values of the absorbed power density ($Q$). Results are shown for four combinations of the initial state (↑,↓) and helicities ($\sigma^\pm$) as indicated at the left side. The HD-AOS is shown in panel corresponding to $Q = 5.8 \times 10^{21}$ W/m³. (b) RE magnetic state after every pulse for $Q = 5.9 \times 10^{21}$ W/m³, showing the appearance of a ring due to the MCD and the switching of the central part. Here pulses with left-handed chirality are applied ($\sigma^+$). The sample side is $\ell = 20$ µm and the laser spot diameter is $d_0 = \ell/2$.

Moreover, the inclusion of the MCD in our eLLB model allows us to explain the experimental observation of rings [28,36,37] which appear after the application of a second laser pulse. This is illustrated in FIG. 5(b) for pulses with $Q = 5.9 \times 10^{21}$ W/m³ and $\tau_L = 50$ fs. The central part of the sample reaches temperatures that lead to HI-AOS, and therefore, its magnetization reverses after each pulse. On the contrary, the ring around of the inverted region is within the HD-AOS regime, and therefore, its local magnetic state (going from *up* to *down*) is only reversed by the first pulse. For the second and subsequent pulses, the ring maintains its *down* state while the inner part changes again to *up* (white). This is repeated every pulse, with the inversion of the central part and the maintenance of black ring in the external shell, as it is clearly seen in even pulses (see snapshots after pulses #2 and #4 in FIG. 5(b)). Note that this ring structure differs from the ones shown in FIG. 2(b) and FIG.3, as they were caused by the inversion of the magnetization around the central demagnetized part under high-power linear pulses ($\sigma = 0$). For circularly polarized pulses ($\sigma^\pm = \pm 1$) the images correspond to alternative switching and the HD-AOS without the central demagnetized (multidomain) state. Again, these results are in good agreement with recent experimental observations (see figures in [28,36,37]).

Instead of the MCD, several other works claim that the observations of the HD-AOS can be ascribed to the ***Inverse Faraday Effect*** (IFE) [9]. Within this formalism, the laser pulse generates an effective out-of-plane magneto-optical field which direction depends on the laser pulse helicity, $\vec{B}_{MO}(\vec{r},t) = \sigma^\pm B_{MO} \eta(r)\epsilon(t)\vec{u}_z$, where $\eta(r) = \exp[-4\ln(2)\, r^2/(2r_0)^2]$ is the spatial field profile, and $\epsilon(t)$ is its temporal profile. Note that the spatial dependence of $\vec{B}_{MO}(\vec{r},t)$ is the same as the one of the absorbed power density. However, according to the literature [9], the so-called magneto-optical field $\vec{B}_{MO}(\vec{r},t)$ has some temporal persistence with respect to the laser pulse, and therefore its temporal profile is different for $t < t_0$ and $t > t_0$: $\epsilon(t < t_0) = \exp[-4\ln(2)(t - t_0)^2/\tau_L^2]$, and $\epsilon(t \geq t_0) = \exp[-4\ln(2)(t - t_0)^2/(\tau_L + \tau_D)^2]$, where $\tau_D$ is the delay time of the $\vec{B}_{MO}(\vec{r}, t)$ with respect to the laser pulse. We have evaluated this IFE scenario by including this field $\vec{B}_{MO}(\vec{r}, t)$ in the effective field of Eq. (4). The results for the same FiM alloy considered up to here (Gd$_x$(FeCo)$_{1-x}$, with $x$=0.25, see SN5), are similar to the ones already presented in FIG. 5 for the MCD considering a maximum magneto-optical field of $B_{MO} = 20$ T with a delay time of $\tau_D = \tau_L$. These IFE results can be seen in FIG. S8 in Supplemental Material Note SN6 [23]. Therefore, we could conclude from this analysis that, from the micromagnetic modeling point of view, both the MCD and the IFE are compatible with experimental observations of the HD-AOS. At this point, it is worth to mention here that in real experiments there is not a clear distinction between MCD and IFE phenomena. Indeed, the modeling of the IFE for absorbing materials can account for absorption phenomena as MCD (see for instance [38,39]). These works suggested that in micromagnetic simulations the IFE could induce a change of the magnetic moment ($\Delta\vec{m}_i$) modifying the initial magnetic moments in the two sublattices of the FiM when submitted to circular polarized laser pulses. We have evaluated in our modeling this alternative manner of studying the role of the IFE by adding such an induced magnetic moment in the eLLB Eq. (4), and compared the results to the case where the IFE is simulated by the magneto-optical field $\vec{B}_{MO}(\vec{r}, t)$ as discussed above. As presented and discussed in Supplemental Material Note SN7 [23], both alternatives ($\vec{B}_{MO}(\vec{r}, t)$ or $\Delta\vec{m}_i$) are equivalent from the simulation point of view. Therefore, in what follows we will simulate the IFE as an effective out-of-plane magneto-optical field.

In order to get a further understanding on the physics of these two mechanisms, either the MCD or the IFE, we have explored the switching probability as a function of $Q$ for laser pulses with fixed duration ($\tau_L = 50$ fs) in two FiM alloys with slightly different composition: Gd$_x$(FeCo)$_{1-x}$ with $x$=0.25 and $x$=0.24 respectively. The corresponding parameters to numerically evaluate these two alloys are given in SN8, and the temperature dependence of the saturation magnetization of each sublattice (RE: Gd; TM: CoFe) are shown in FIG. 6 (a) and (c) respectively. Note that magnetization compensation temperature at which the net magnetization of the sample vanishes ($T_M$) is above and below room temperature for $x$=0.25 and $x$=0.24 respectively. In other words, the FiM sample is dominated by the RE (TM) at $T = 300$ K for $x$=0.25 ($x$=0.24) compositions. The MCD and IFE parameters remain fixed as indicated above (MCD:

$k \sim 2\%$; IFE: $B_{MO} = 20$ T, $\tau_D = \tau_L$). The two possible initial states prior the laser pulse ($m_z^{TM}: (\uparrow, \downarrow)$), and the three laser polarizations (linear: $\sigma = 0$, and circular $\sigma^\pm = \pm 1$) were evaluated. The switching probability was computed by evaluating ten different stochastic realizations for each for each $Q$, and the results are presented in FIG. 6(b) and (d) for $x$=0.25 and $x$=0.24 respectively. As in experimental observations [9], the HD-AOS takes place only in a narrow range of $Q$ around the threshold value of $Q$ at which the switching probability abruptly changes from 0 to 1 under linear polarized pulses (black dots in FIG. 6(b) and (d)).

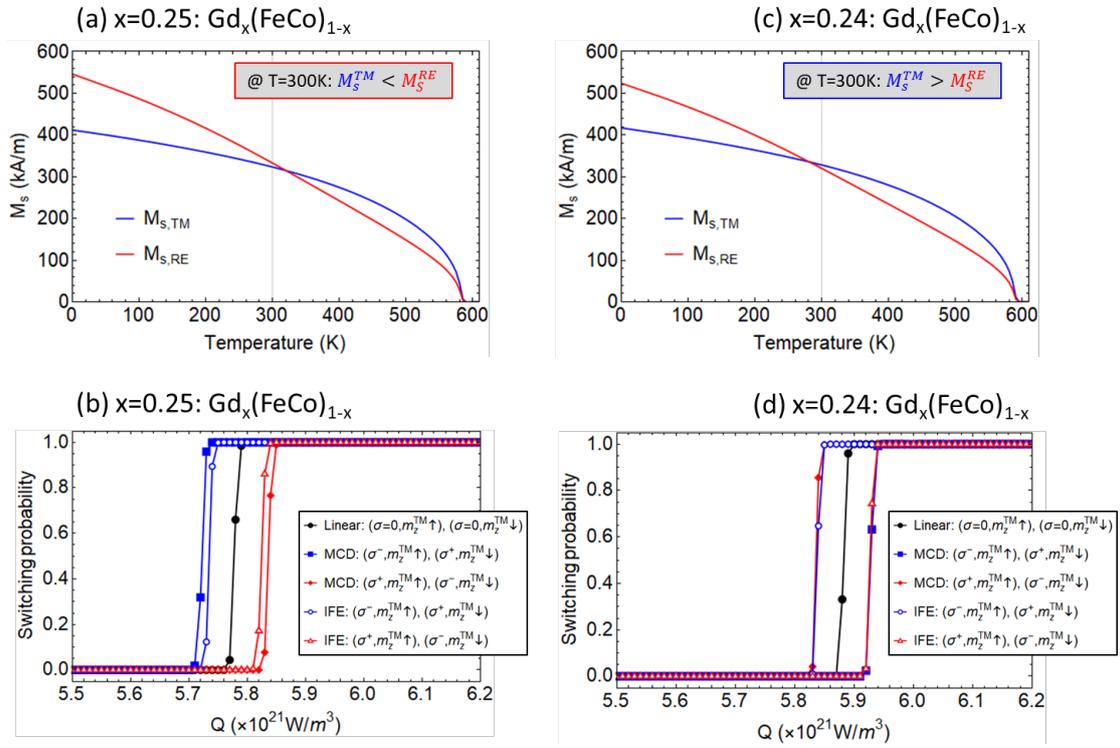

FIG. 6. Temperature dependence of the spontaneous magnetization of each sublattice (RE:Gd; TM:CoFe) of the FiM alloy (Gd$_x$(CoFe)$_{1-x}$) for two different compositions: (a) $x$=0.25 and (c) $x$=0.24. The vertical grey line indicates the initial room temperature prior the laser pulse ($T = 300$ K). Probability of switching as a function of the absorbed power density ($Q$) for a laser pulse of $\tau_L = 50$ fs for different combinations of the initial state ($m_z^{TM}: (\uparrow, \downarrow)$) and the polarization (linear: $\sigma = 0$ (black dots), and circular $\sigma^\pm = \pm 1$) of the laser pulse as indicated in the legend and in the main text: (b) corresponds to $x$=0.25 and (d) to $x$=0.24. MCD results are shown by solid dots, whereas IFE results are presented by open symbols. Lines are guide to the eyes.

For $x$=0.25, as all results presented up to here, the FiM is dominated by the RE:Gd at room temperature: $M_s^{TM} < M_s^{RE}$ at $T = 300$K, see FIG. 6(a). In this case, the switching requires less $Q$ with circular polarization ($\sigma^\pm$) with respect to the linearly polarized case ($\sigma = 0$) for two different combinations of the circular laser polarization and the initial state of the FiM: ($\sigma^-, m_z^{TM} \uparrow$) and ($\sigma^+, m_z^{TM} \downarrow$). This happens for both MCD (solid symbols) and IFE scenarios (open symbols) as it is shown in FIG. 6(b). Note that the initial state in the TM is the opposite to the RE: $m_z^{TM} \uparrow$ (*up*) corresponds to $m_z^{RE} \downarrow$ (*down*) and vice, and the AOS is independent on the initial state for linear polarization (HI-AOS), whereas under circular polarized laser pulses the switching is helicity-dependent (HD-AOS). For the rest of combinations, either ($\sigma^+, m_z^{TM} \uparrow$) or ($\sigma^-, m_z^{TM} \downarrow$), a higher $Q$ is needed to achieve 100% of switching probability with respect to the linear polarized laser pulse, and again for *x*=0.25 both MCD and IFE scenarios result in similar behavior of the switching probability (FIG. 6(b)).

Remarkably, the MCD and IFE results are qualitatively different when the composition is slightly modified to *x*=0.24, where the FiM becomes dominated by the TM:CoFe at room temperature: $M_s^{TM} > M_s^{RE}$ at $T = 300K$, see FIG. 6(c). In this case (*x*=0.24), the results in the IFE scenario are qualitatively similar to the ones already obtained for *x*=0.25: the HD-AOS occurs with small $Q$ with respect to the linearly polarization case for ($\sigma^-, m_z^{TM} \uparrow$) and ($\sigma^+, m_z^{TM} \downarrow$) (open blue symbols in FIG. 6(d)), and it requires high $Q$ for the two other combinations (($\sigma^+, m_z^{TM} \uparrow$), ($\sigma^-, m_z^{TM} \downarrow$), open red symbols in FIG. 6(d)). However, for this concentration (*x*=0.24), the results in the MCD scenario (see solid symbols in FIG. 6(d)) are opposite as for *x*=0.25, and also opposite to the ones obtained in the IFE scenario.

These results can be understood as follows. In the MCD scenario, the HD-AOS depends on the net initial magnetization at room temperature ($m_N = M_s^{TM} m_z^{TM} + M_s^{RE} m_z^{RE}$, with $m_z^{TM} = \pm 1$ and $m_z^{RE} = \mp 1$) and on the laser helicity ($\sigma^\pm = \pm 1$): if initially $m_N > 0$, a laser pulse with $\sigma^+ = +1$ promotes the reversal, and this happens either for $m_z^{TM} = -1$ ($m_z^{RE} = +1$) when *x*=0.25 because $M_s^{RE} > M_s^{TM}$ at $T = 300K$, or for $m_z^{TM} = +1$ ($m_z^{RE} = -1$) if *x*=0.24 because now $M_s^{TM} > M_s^{RE}$ at $T = 300K$. On the other hand, the HD-AOS within the IFE scenario is essentially determined by the dominant sublattice magnetization just below the Curie threshold ($T \lesssim T_C$), due to the persistence of the magneto-optical field $\vec{B}_{MO}(\vec{r}, t)$ when the laser pulse has already finished. Note that for both concentrations $M_s^{TM} > M_s^{RE}$ for $T \lesssim T_C$. Indeed, during the cooling down after the pulse, the magneto-optical field $\vec{B}_{MO}(\vec{r}, t) \propto \sigma^\pm \vec{u}_z$ promotes *up* or *down* magnetic state for the TM for $\sigma^+$ and $\sigma^-$ respectively, and therefore, if $m_z^{TM}$ is initially *up* ($m_z^{TM} = +1$), $\vec{B}_{MO}(\vec{r}, t)$ promotes the reversal for $\sigma^-$ and vice. Our analysis suggests a set of experiments which could help to elucidate the physical mechanism behind these HD-AOS, just by evaluating the switching probability as function of the initial state and of the laser pulse helicity for two different concentrations *x*, one resulting in a TM-dominated and other in a RE-dominated FiM alloy at room temperature. Another alternative could be to use a single FiM with a given composition, and working with a cryostat to fix different initial temperatures below and above the magnetization compensation temperature. Similar results to ones obtained by changing the composition *x* for a fixed temperature of the thermal bath are also predicted by our simulations when it is the temperature of the thermal bath what is varied for a given composition (see Fig. S10 in Supplemental Material Note SN9 [23]): both IFE and MCD scenarios give similar results below compensation and opposite above it. These theoretical predictions on the HD-AOS could be checked by experiments, which all together would allow us to shed light on the real scope of these two mechanisms.

## IV. CONCLUSIONS

As summary, the extension of the two sublattice Landau-Lifshitz-Bloch equation with the angular momentum non-equilibrium exchange is proven to be a powerful tool to study ultrafast AOS switching in ferrimagnetic alloys. The formalism here developed reproduces the atomistic spin dynamics results for small samples at the nanoscale, while it opens the possibility to numerically study realistic extended micro-size systems, with dimensions comparable to the experimental ones. The deterministic single-shot switching and the demagnetization at high power regime are found to be in very good agreement with the experimental observations of Helicity-Independent AOS under linearly polarized laser pulses. The phase diagram combined with the thermal analysis allowed us to determine and predict the size of the inverted regions as depending the absorbed power and duration of the laser pulse. Moreover, we have also explored and reproduced experimental observations for the Helicity-Dependent AOS within the two physical mechanisms suggested in the literature: Magnetic Circular Dichroism and Inverse Faraday Effect. According to the Magnetic Circular Dichroism, the absorbed power by the FiM depends on the laser helicity under circularly polarized pulses, and our model also predicts the main features of the Helicity-Dependent AOS measurements. Indeed, both the Helicity-Dependent AOS and the appearance of rings around the circularly polarized laser beam appear naturally in our simulations. Additionally, similar results of the HD-AOS switching were also obtained in Inverse Faraday Effect scenario, where the circular polarization has been suggested to generate a persistent magneto-optical field promoting the switching for proper combinations of initial magnetic state and laser pulse helicity. By exploring FiM samples with different compositions resulting in TM-dominated or in a RE-dominated FiM alloy at room temperature, we have found a difference between the predictions of the IFE and the MCD scenarios. These results could be tested by performing the corresponding experiments, and consequently helping together to elucidate the true basis of such HD-AOS processes. Therefore, our methods will be useful to understand recent and future experiments on AOS, and also to the develop novel recording devices where the information can be manipulated by optical means in an ultra-fast fashion.

## ACKNOWLEDGMENTS

This work was supported by projects MAT2017-87072-C4-1-P funded by Ministerio de Educacion y Ciencia and PID2020117024GB-C41 funded by Ministerio de Ciencia e Innovacion, both from the Spanish government, projects No. SA299P18 and SA114P20 from Consejeria de Educacion of Junta de Castilla y León, and project MagnEFi, Grant Agreement 860060, (H2020-MSCA-ITN-2019) funded by the European Commission. UA gratefully acknowledges support by the Deutsche Forschungsgemeinschaft through SFB/TRR 227 "Ultrafast Spin Dynamics", Project A08.

**Supplemental Material: Realistic micromagnetic description of all-optical ultrafast switching processes in ferrimagnetic alloys**

V. Raposo[1,*], F. García-Sánchez[1], U. Atxitia[2], and E. Martínez[1,+]

1. *Applied Physics Department, University of Salamanca.*
2. *Dahlem Center for Complex Quantum Systems and Fachbereich Physik.*

*,+ Corresponding authors: victor@usal.es, edumartinez@usal.es

**CONTENT:**

- **Supplemental Note SN1.** *Atomistic Spin Dynamics (ASD) model*
- **Supplemental Note SN2.** *Micromagnetic LLB models: conventional (LLB) and extended (eLLB) cases*
- **Supplemental Note SN3.** *Comparison between atomistic, conventional-LLB and extended-LLB models*
- **Supplemental Note SN4.** *Phase diagram in terms of the fluence and the pulse duration*
- **Supplemental Note SN5.** *Helicity-Dependent AOS (HD-AOS) and Magnetic Circular Dichroism (MCD)*
- **Supplemental Note SN6.** *Helicity-Dependent AOS (HD-AOS) and Inverse Faraday Effect (IFE)*
- **Supplemental Note SN7.** I*nverse Faraday Effect: magneto-optical field or induced magnetic moment*
- **Supplemental note SN8.** *Material inputs for two different compositions*
- **Supplemental note SN9:** *Helicity-Dependent All Optical Switching: MCD & IFE for different compositions and initial temperatures*

## SN1. *Atomistic Spin Dynamics (ASD) model*

With the growing power of modern computers, numerical modeling has become an essential tool for scientific research, especially when handling systems as complex as magnetic devices. For magnetic modeling, there are several methodological choices in terms of dimensions and time scales. In this section we review the bases of the *Atomistic Spin Dynamic* (ASD) model used to study the magnetization dynamics in small nano-size ferrimagnetic FiM samples submitted to ultra-short of laser pulses. The physical basis of the ASD model is the localization of unpaired electrons to atomic sites, leading to an effective local atomistic magnetic moment at site $i$, $(\vec{\mu}_i)$, which is treated as a classical spin of fixed length. The FiM alloys are composed of two sublattices: rare-earth (RE) and transition-metal (TM). Typical examples of these FiM alloys are GdCo, GdFe, GdFeCo, TbFeCo, etc. We consider that the ordered TM alloy is represented by the fcc-type lattice, and to simulate the amorphous character of the TM-RE alloy, $x_{RE} \cdot 100\%$ lattice sites are substituted randomly with RE magnetic moments. An example of a typical atomistic arrangement is shown in Fig. S1 for a $Gd_x(FeCo)_{1-x}$ with $x = x_{RE} = 0.25$ being the relative composition of the sublattices ($x = x_{RE}$, and $x_{TM} = 1 - x_{RE}$). Each atom, either RE (Gd) or TM (FeCo), has its local atomistic magnetic moment $\vec{\mu}_i$ at site $i$.

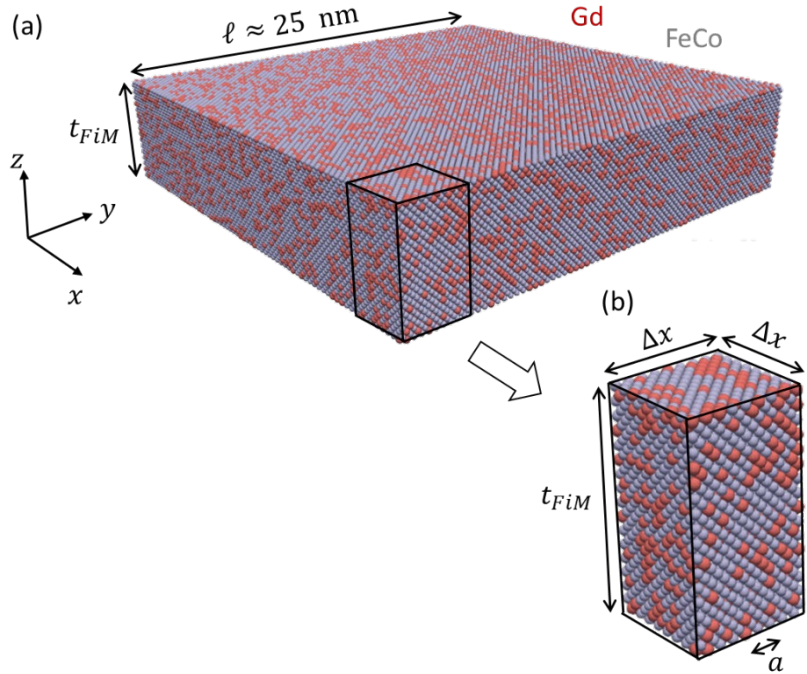

FIG. S1. Atomistic scheme showing a FiM $Gd_x(FeCo)_{1-x}$ alloy consisting on two sublattices of RE (Gd, red) and TM (CoFe, grey) magnetic moments, with $x$ representing the relative composition of each sublattice. The sample dimensions along the Cartesian coordinate's directions are $(\ell \times \ell \times t_{FiM})$ with $t_{FiM} = 5.6$ nm. Atomistic simulations are restricted to samples at the nanoscale ($\ell \approx 25$ nm). The dimensions of an elementary computational cell ($\Delta x = \Delta y = 3$ nm) within the micromagnetic approach in shown at the bottom, along with the indication of the lattice constant ($a = 0.32$ nm) for comparison.

The basis of ASD model (see for instance [1,2] and references therein) is a classical spin Hamiltonian based on the Heisenberg exchange formalism. The spin Hamiltonian $\mathcal{H}$ typically has the form:

$$\mathcal{H} = \mathcal{H}_{exc} + \mathcal{H}_{ani} + \mathcal{H}_{dip} + \mathcal{H}_{app} \tag{eS1}$$

with the terms on the RHS representing respectively exchange, anisotropy, dipolar and Zeeman terms. The exchange energy for a system of interacting atomic moments is given by the expression

$$\mathcal{H}_{exc} = -\sum_{i \neq j} J_{ij} \vec{S}_i \cdot \vec{S}_j \tag{eS2}$$

where $J_{ij}$ is the exchange interaction between atomic sites $i$ and $j$, $\vec{S}_i$ is a unit vector denoting the local spin moment direction ($\vec{S}_i = \vec{\mu}_i/\mu_{si}$ with $\mu_{si} = |\vec{\mu}_{si}|$) and $\vec{S}_j$ is the spin moment direction of neighboring atoms. The sum in Eq. (eS2) is truncated to nearest neighbors only. As the FiM alloy is formed by two sublattices of magnetic moments, we can split the exchange interaction ($J_{ij}$) between ferromagnetic intra-lattice ($J_{ij} > 0$) and antiferromagnetic inter-lattice ($J_{ij} < 0$) exchange interactions. In what follows, we adopt the notation of $J_{TM-TM} > 0$, $J_{RE-RE} > 0$ for intra-lattice interactions, and $J_{TM-RE} < 0$ for the inter-lattice interactions.

The second term in Eq. (eS2) is the magnetic anisotropy. Here, a standard uniaxial anisotropy along the easy-axis ($z$, out-of-plane direction) is considered,

$$\mathcal{H}_{ani} = -d_u \sum_i (\vec{S}_i \cdot \vec{u}_K)^2 \tag{eS3}$$

where $d_u$ is the anisotropy energy per atom and $\vec{u}_K = \vec{u}_z$ is the unit vector denoting the preferred direction. The last two terms in Eq. (eS1) account for the dipolar ($\mathcal{H}_{dip}$) and external applied field ($\mathcal{H}_{app}$) contributions. Since the demagnetizing field is usually relatively small in FiMs samples, this term is generally ignored here. No external field is applied in the present work.

The equilibrium state of the FiM sample can be obtained by minimizing of the total energy. Here, it consists of the two antiferromagnetic coupled sublattices aligned along the easy-axis in anti-parallelly. Its dynamics response is governed by the Langevin-Laudau-Lifshitz-Gilbert equation for each atomistic moment ($\vec{S}_i$) [1,2]:

$$\frac{\partial \vec{S}_i}{\partial t} = -\frac{\gamma_0}{(1+\lambda^2)} \{\vec{S}_i \times (\vec{H}_i + \vec{H}_{th,i}) + \lambda \vec{S}_i \times [\vec{S}_i \times (\vec{H}_i + \vec{H}_{th,i})]\} \tag{eS4}$$

where $\lambda = 0.02$ is the Gilbert damping parameter and $\gamma_0 = 2.21 \times 10^5$ m/(A·s) the gyromagnetic ratio. $\vec{H}_i$ is the local effective magnetic field obtained from the spin Hamiltonian as

$$\mu_0 \vec{H}_i = -\frac{1}{\mu_{si}} \frac{\partial \mathcal{H}}{\partial \vec{S}_i} \tag{eS5}$$

and $\vec{H}_{th,i}$ is the stochastic thermal field is given by:

$$\mu_0 \vec{H}_{th,i} = \vec{\Gamma}_i(t) \sqrt{\frac{2\lambda k_B T}{\gamma_0 \mu_{si} \Delta t}} \quad \text{(eS6)}$$

with $k_B$ is the Boltzmann constant, $T$ the temperature, $\Delta t$ the integration time step. $\vec{\Gamma}_i(t)$ is a local vector whose components are Gaussian-distributed white-noise random numbers with zero mean value.

Under a laser pulse, the FiM sample absorbs energy and its temperature changes in time. The temperature $T$ that enters in Eq. (eS6) is the temperature of the electronic subsystem ($T \equiv T_e$), which is coupled to the lattice subsystem temperature ($T_l$) and to the laser pulse as given by the Two Temperature Model (TTM). Details of the power absorbed by sample as due to the laser pulse and the Eqs. (2)-(3) of the TTM were already discussed in the main text. Conventional TTM values were adopted [3–5]: $C_e = 1.8 \times 10^5$ J/(m³K) at $T_R = 300$ K, $C_l = 3.8 \times 10^6$ J/(m³K), $k_e = 91$ W/(m K), $g_{el} = 7 \times 10^{17}$ W/(m³K) and $\tau_D = 10$ ps.

Eq. (eS4) [or Eq. (1) in the main text] is numerically solved coupled to the TTM Eqs. ((2)-(3) in the main text) with a home-made solver using a 4$^{th}$-order Runge-Kutta scheme with $\Delta t = 0.1$ fs. Typical Gd$_{0.25}$(CoFe)$_{0.75}$ parameters were considered [2]: intra-lattice exchange energies: intra-lattice exchange energies $J_{TM-TM} = 3.58 \times 10^{-21}$ J, $J_{RE-RE} = 1.44 \times 10^{-21}$ J, inter-lattice energy $J_{TM-RE} = -1.25 \times 10^{-21}$ J, magnitude of the local magnetic moments: $\mu_{TM} = 1.92\ \mu_B$, $\mu_{RE} = 7.63\ \mu_B$, and anisotropy energy of $d_u = 8.07 \times 10^{-24}$ J. The FiM sample dimensions for atomistic simulations are $\ell \times \ell \times t_{FiM}$ (see FIG. S1) with $\ell \simeq 25$ nm and $t_{FiM} = 5.6$ nm, and magnetic moment of the two sublattices are randomly generated with 25% and 75% of RE:Gd and TM:(FeCo) respectively (FIG. S1). The lattice constant is $a = 0.35$ nm. TTM parameter The power absorbed by the laser pulse is assumed to be uniform over the nano-size FiM sample considered in the main text for atomistic simulations (ASD).

## SN2. *Micromagnetic LLB models: conventional (LLB) and extended (eLLB) cases*

As it was already stated, due to memory and computation limitations, atomistic simulations are restricted to small samples at the nano-scale ($\ell \lesssim 100$ nm). However, experimental studies on All Optical Switching are typical carried out in extended samples with several micrometers in length ($\ell \simeq 100$ μm). Also, the size of conventional laser spot (FWHM) are usually at the micro-scale ($d_0 \sim 1 - 10$ μm). It is worthy to provide an estimation of the maximum size that could be numerically managed in atomistic simulations (ASD). For instance, in the present work we used one of the most powerful Graphics Processing Units (GPUs), model RTX3090 (24GB), and performing atomistic simulations of FiM samples of $\ell \times \ell \times t_{FiM}$ with $\ell \simeq 150$ nm and $t_{FiM} = 5.6$ nm ($a \sim 0.35$ nm) will be already out of the computational power of such GPU. Therefore, in order to explain experimental observations another coarse-grained formalism is needed. Here, we adopt the mesoscopic description (micromagnetic model) which assumes that the magnetization of each sublattice is a continuous vector field: $\vec{m}_i = \vec{m}_i(\vec{r}, t)$, where here the sub-index $i$ refers to the local magnetic magnetization of each sublattice, $i$:RE,TM. The FiM sample, with dimension $\ell \times \ell \times t_{FiM}$, is discretized using a 2D finite differences scheme, using computational cells of volume $\Delta V = \Delta x^2 \, t_{FiM}$. The continuous description ($\Delta x \gg a$) is justified by the short-range of the exchange interaction, which promotes the ferromagnetic and the antiferromagnetic orders for intra- and inter-lattices interactions respectively. At the same time, in order to resolve magnetic patterns such as domain walls, the cell size $\Delta x$ must be smaller than the characteristic length scale along which the magnetization varies significantly (for instance, the so-called domain wall width, $\delta \gg \Delta x$). Therefore, within the micromagnetic approach, at each cell location ($\vec{r}$) there are two magnetic spins representing the magnetization of the two sublattices of the FiM, $\vec{m}_i = \vec{m}_i(\vec{r},t) = \vec{M}_i(\vec{r},t)/M_{si}(T)$ where $\vec{M}_i(\vec{r},t)$ is the local magnetization of sublattice $i$ in units of A/m, and $M_{si}(T)$ is the corresponding spontaneous magnetization at temperature $T$. FIG. S2 shows these ideas, along with different spatial scales of the ASD and micromagnetic model.

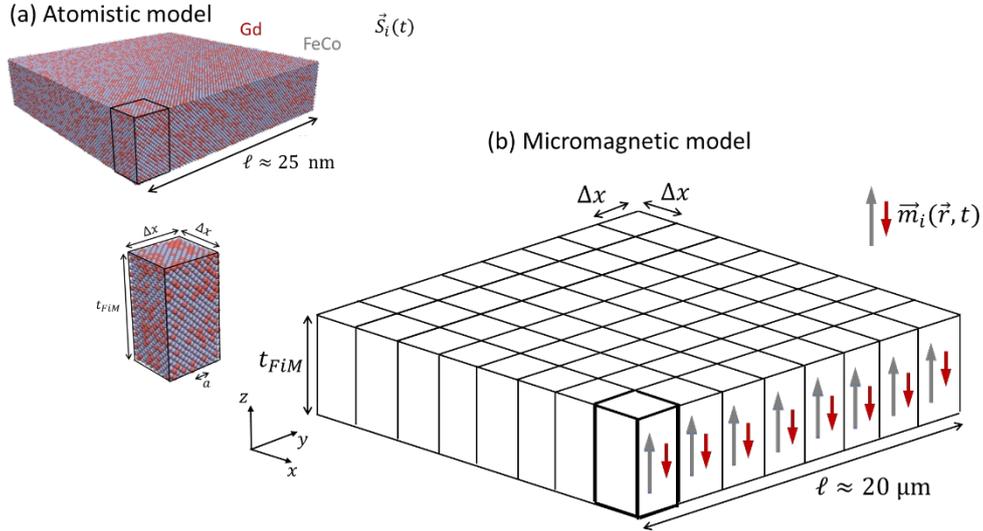

FIG. S2. (a) Arrange of atoms for atomistic simulations, limited to nanoscale samples, $\ell \approx 25$ nm. (b) Micromagnetic discretization scheme for extended samples at the microscale, $\ell \approx 20$ µm. The FiM sample is discretized using a 2D finite differences scheme using computational cells of volume $V = \Delta x^2 \, t_{FiM}$. Each cell contains two micro-magnetic moments, one for each sublattice, and its size would include thousands of atomistic magnetic moments ($\vec{S}_i$). The magnetization of each sublattice is a continuous vector function ($\vec{m}_i = \vec{m}_i(\vec{r},t)$) over the FiM sample.

In order to overcome the ASD limitations here we adopt an extended continuous micromagnetic model that describes the temporal evolution of the reduced local magnetization $\vec{m}_i(\vec{r},t)$ of each sublattice $i$:RE,TM based on the stochastic Landau-Lifshitz-Bloch (LLB) equation [1,6]

$$\frac{\partial \vec{m}_i}{\partial t} = -\gamma'_{0i}(\vec{m}_i \times \vec{H}_i) +$$

$$-\frac{\gamma'_{0i}\alpha_i^\perp}{m_i^2}\vec{m}_i \times [\vec{m}_i \times (\vec{H}_i + \vec{\xi}_i^\perp)] - \quad \text{(eS7)}$$

$$+\frac{\gamma'_{0i}\alpha_i^\parallel}{m_i^2}(\vec{m}_i \cdot \vec{H}_i)\vec{m}_i + \vec{\xi}_i^\parallel$$

where $\gamma'_{0i} = \gamma_{0i}/(1 + \lambda_i^2)$ is the reduced gyromagnetic ratio, which is defined via the coupling parameter $\lambda_i$ of sublattice $i$ to the heat bath. $\alpha_i^\parallel$ and $\alpha_i^\perp$ are the longitudinal and perpendicular damping parameters. $\vec{H}_i = \vec{H}_i(\vec{r}, t)$ is the local effective field at location $\vec{r}$ acting on sublattice magnetic moment $i$, and $\vec{\xi}_i^\parallel$ and $\vec{\xi}_i^\perp$ are the longitudinal and perpendicular stochastic thermal fields.

For $T < T_C$, the damping constants for sublattice $i$ are

$$\alpha_i^\parallel = 2\lambda_i k_B T \frac{m_{e,i}}{J_{0,ii}m_{e,i} + |J_{0,ij}|m_{e,i}} \quad \text{(eS8)}$$

$$\alpha_i^\perp = \lambda_i \left(1 - k_B T \frac{m_{e,i}}{J_{0,ii}m_{e,i} + |J_{0,ij}|m_{e,j}}\right) \quad \text{(eS9)}$$

where $m_{e,i}$ and $m_{e,j}$ are the equilibrium magnetization in sublattices $TM$ and $RE$ respectively. These values are obtained from the mean field approximation (see Ref. [1,6]). The exchange energies became $J_{0,ii} = zx_i J_{i-i}$, $J_{0,ij} = zx_j J_{i-j}$, with $z$ the number of nearest neighbors and $J_{i-i}$ and $J_{i-j}$ the atomistic exchange energy between species. $x_i$ is the concentration of sublattice $i$. Above the Curie temperature, $T > T_C$, the damping parameters are

$$\alpha_i^\parallel = \alpha_i^\perp = \frac{2\lambda_i T}{3T_C} \quad \text{(eS10)}$$

The effective field consists on several contributions:

$$\vec{H}_i = \vec{H}_{ani,i} + \vec{H}_{ex,i} + \vec{H}_i^\parallel \quad \text{(eS11)}$$

where $\vec{H}_{ani,i}$ and $\vec{H}_{ex,i}$ are the anisotropy and exchange fields given by

$$\vec{H}_{ani,i} = \frac{2K_i}{\mu_0 \mu_i}\vec{m}_i \cdot \vec{u}_K \quad \text{(eS12)}$$

$$\vec{H}_{ex,i} = \frac{2A_{ex,i}}{\mu_0 M_{si}}(\nabla^2 \vec{m}_i) - \frac{J_{0,ij}}{\mu_0 \mu_i m_i^2}[\vec{m}_i \times (\vec{m}_i \times \vec{m}_j)] \quad \text{(eS13)}$$

where the first term in $\vec{H}_{ex,i}$ is the continuous exchange between neighbors' cells, and the second one accounts for the inter-lattice contribution. Finally, the term $\vec{H}_i^\parallel$ is a field computed differently below $T_C$,

$$\vec{H}_i^\parallel(T < T_C) = -\frac{1}{\mu_0}\left(\frac{1}{\tilde{\chi}_i^\parallel} + \frac{|J_{0,ij}|\tilde{\chi}_j^\parallel}{\mu_i \tilde{\chi}_i^\parallel}\right)\frac{\delta m_i}{m_{e,i}}\vec{m}_i + \frac{1}{\mu_0}\frac{|J_{0,ij}|\delta\tau_{B,i}}{\mu_i m_{e,i}}\vec{m}_i \quad \text{(eS14)}$$

and above $T_C$

$$\vec{H}_i^{\parallel}(T \geq T_C) = -\frac{1}{\mu_0}\left(\frac{1}{\tilde{\chi}_i^{\parallel}} + \frac{|J_{0,ij}|\tilde{\chi}_j^{\parallel}}{\mu_i \tilde{\chi}_i^{\parallel}}\right)\vec{m}_i + \frac{1}{\mu_0}\frac{|J_{0,ij}|}{\mu_i}\frac{\tau_{B,i}}{m_i}\vec{m}_i \quad (eS15)$$

with $\tau_{B,i} = |\vec{m}_i \cdot \vec{m}_j|/m_i$, $\delta m_i = m_i - m_{e,i}$ and $\delta\tau_{B,i} = \tau_{B,i} - \tau_{e,B,i}$. The longitudinal susceptibilities $\tilde{\chi}_i^{\parallel}$ are calculated from mean field approach [1]. Below the Curie temperature ($T < T_C$):

$$\tilde{\chi}_i^{\parallel}(T < T_C)$$
$$= \frac{\mu_j \mathcal{L}_i'(\zeta_i)|J_{0,ij}|\mathcal{L}_j'(\zeta_j) + \mu_j \mathcal{L}_i'(\zeta_i)[k_B T - J_{0,jj}\mathcal{L}_j'(\zeta_j)]}{[k_B T - J_{0,ii}\mathcal{L}_i'(\zeta_i)][k_B T - J_{0,jj}\mathcal{L}_j'(\zeta_j)] - |J_{0,ji}|\mathcal{L}_i'(\zeta_i)|J_{0,ij}|\mathcal{L}_j'(\zeta_j)} \quad (eS16)$$

where $\mathcal{L}_i$ is the Langevin function with argument $\zeta_i = J_{0,ij}m_i + \frac{|J_{0,ij}|m_j}{k_B T}$ and $\mathcal{L}_i'$ is the derivative respect $\zeta_i$. Above the Curie temperature ($T > T_C$): [7]

$$\tilde{\chi}_i^{\parallel}(T \geq T_C) = \frac{\mu_j|J_{0,ij}| + \mu_j(3k_B T - J_{0,jj})}{\left((3k_B T - J_{0,ii})(3k_B T - J_{0,jj}) - |J_{0,ij}||J_{0,ji}|\right)} \quad (eS17)$$

Stochastic fluctuations due to temperature are introduced through thermal fields $\vec{\xi}_i^{\parallel}$ and $\vec{\xi}_i^{\perp}$. These fields are time and space uncorrelated, and they are generated from white noise random numbers with zero mean field and variance given by [6]:

$$\langle \xi_{i,\alpha}^{\eta}(\vec{r},t)\xi_{i,\beta}^{\eta}(\vec{r}',t')\rangle = 2D_i^{\eta}\delta_{\alpha\beta}\delta(\vec{r}-\vec{r}')\delta(t-t') \quad (eS18)$$

where $\alpha,\beta$ are the cartesian components of the stochastic thermal fields, $i$ denotes sublattice $TM$ or $RE$ and $\eta$: $\parallel,\perp$ represents parallel or perpendicular field components. The diffusions constants are given by:

$$D_i^{\perp} = \frac{(\alpha_i^{\perp} - \alpha_i^{\parallel})a^3 k_B T}{(\alpha_i^{\perp})^2 \gamma_{0i}'\mu_0 n_{at}x_i\mu_i \Delta V} \quad (eS19)$$

$$D_i^{\parallel} = \frac{\alpha_i^{\parallel}\gamma_{0i}'a^3 k_B T}{n_{at}x_i\mu_0\mu_i \Delta V} \quad (eS20)$$

with $\Delta V = \Delta x^2 t_{FiM}$ the discretization volume, $a$ is the lattice constant and $n_{at}$ is the number of atoms per unit cell.

As it will be shown in Supplemental Note N3(b), we have confirmed that the conventional LLB Eq. (eS7) (or Eq. (4) in the main text) [1,6,8] is not able to fully reproduce the atomistic model predictions of some all optical switching processes in FiM systems. Indeed, atomistic simulations show that the sudden change in the temperature due to the laser pulse results in transient non-equilibrium states where a transfer of angular momentum between sublattices takes place. This transference is not fully reproduced by the conventional LLB Eq. (eS7). In order to overcome this limitation, in the present work we extend the LLB Eq. (eS7) by adding an additional term to the RHS that accounts of the angular momentum transfer between lattices in non-equilibrium states. This non-equilibrium torque is given

$$\vec{\tau}_i^{NE} = \gamma'_{0i} \lambda_{ex} \alpha_i^{\parallel} \frac{x_i \mu_i m_i + x_j \mu_j m_j}{\mu_i m_i \mu_j m_j} (\mu_i \vec{H}_i^{\parallel} - \mu_j \vec{H}_j^{\parallel}) \tag{eS21}$$

where and $\vec{H}_i^{\parallel}$ and $\vec{H}_j^{\parallel}$ are the longitudinal effective fields for each lattice $i$:RE,TM [6], $x_i \equiv x$ and $x_j = 1 - x_i \equiv 1 - x$ are the concentrations of each specimen, and $\lambda_{ex}$ is a parameter representing the exchange relaxation rate [9]. By including Eq. (eS21) in the RHS of Eq. (eS7), and numerically solving it coupled to TTM Eqs. (2)-(3), we can provide a realistic description of the magnetization dynamics in FiM systems under ultra-short laser pulses. In what follows, and in order to distinguish it from the *conventional LLB model* (LLB, $\vec{\tau}_i^{NE} = 0$), we refer to this formalism as the *extended micromagnetic LLB model* (eLLB, $\vec{\tau}_i^{NE} \neq 0$), Note that the additional torque has not been previously included in the LLB model, and therefore, here we name it here as the extended LLB model (eLLB). A comparison between atomistic (ASD) predictions and the results without (conventional LLB) and with (extended LLB, eLLB) the additional torque $\vec{\tau}_i^{NE}$ is shown in FIG. S4, discussed in the next section.

In the present work, FiM samples of $\ell \times \ell \times t_{FiM}$ are discretized using a 2D finite differences scheme, using computational cells of volume $\Delta V = \Delta x^2 \, t_{FiM}$. Typical material parameters for Gd$_{0.25}$(FeCo)$_{0.75}$ were adopted: $J_{0,TM-TM} = 2.59 \times 10^{-21}$ J, $J_{0,RE-RE} = 1.35 \times 10^{-21}$ J, $J_{0,TM-RE} = -1.13 \times 10^{-21}$ J. $\lambda_{TM} = \lambda_{RE} = 0.02$. Micromagnetic parameters needed to solve the LLB Eq. can be obtained from atomistic inputs as follows [2]: the spontaneous magnetization at zero temperature of each sublattice is $M_{si}(0) = \mu_i x_i / (p_f \cdot a^3)$, where $p_f$ is a packing factor which depends on the crystalline structure ($p_f = 0.74$ for fcc, [2]): $M_{S,TM}(0) = 0.41$ MA/m, $M_{S,RE}(0) = 0.55$ MA/m. The continuous exchange stiffness constant [10] is $A_{ex,i} = n\epsilon x_i^2 J_i / (2a)$ with $n = 2$ being the number of atoms per unit cell for fcc, $\epsilon = 0.79$ the spin wave mean field correction value also for FCC, and $x_i$ the concentration of sublattice $i$ ($x_j = 1 - x_i$): $A_{ex,TM} = 3.2$ pJ/m, $A_{ex,RE} = 0.19$ pJ/m. The perpendicular anisotropy parameter is $K_i = d_u x_i / (p_f \cdot a^3)$, see [2], so $K_{u,TM} = 1.87$ MJ/m$^3$, $K_{u,RE} = 0.62$ MJ/m$^3$. For the presented results we take $\lambda_{ex} = 0.013$ for the exchange relaxation rate. Within the micromagnetic model, the laser-induced magnetization dynamics is evaluated by numerically solving Eq. (eS7) coupled to the TTM Eqs. (2)-(3) using the Heun algorithm with $\Delta t = 1$ fs with cell sizes of $\Delta x = 3$ nm. This was done by implementing the models in a home-made CUDA-based code, which was run a RTX3090 GPU. It was also checked in several tests that decreasing the cell size to $\Delta x = 1$ nm does not change the results.

### SN3. *Comparison between atomistic, conventional LLB and extended LLB models*

(a) In order to validate the extended LLB model (eLLB, $\vec{\tau}_i^{NE} \neq 0$), we have firstly compared the predictions of the conventional LLB model (LLB, $\vec{\tau}_i^{NE} = 0$) with the atomistic results (ASD) by computing for the equilibrium magnetization of each sublattice ($i$:RE,TM) as a function of the temperature of the thermal bath (which coincides with the electronic temperature). This study was carried out in a small FiM sample with $\ell \approx 25$ nm, and the results are shown in FIG. S3. Both ASD and eLLB models lead to similar results.

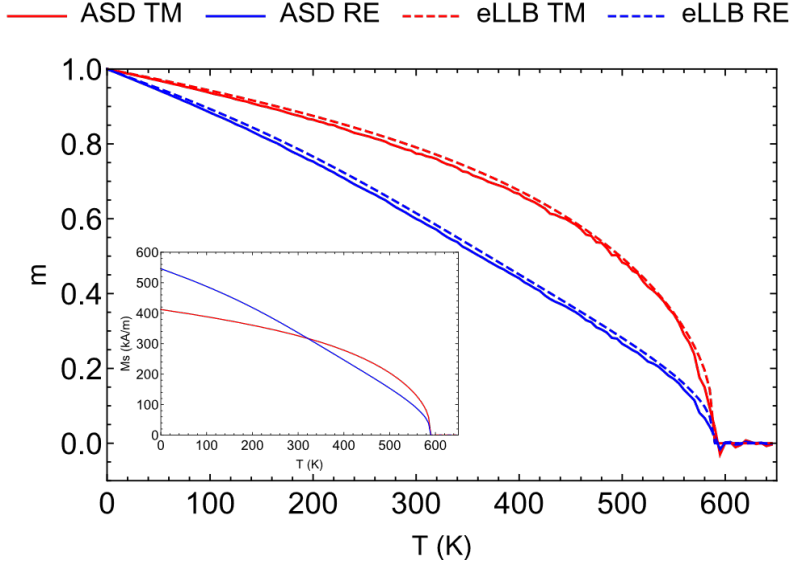

FIG. S3. Temperature dependence of the reduced equilibrium magnetization ($m$) of the two sublattices. Solid (dashed) lines corresponds to ASD (eLLB) results. Blue and red color correspond to the RE and TM sublattices respectively. The inset shows the corresponding dependence of the spontaneous magnetization value of each sublattice ($M_{si}(T)$ vs $T$).

**(b)** In the main text and also in SN2, we claimed that the conventional LLB model (LLB, $\vec{\tau}_{NE,i} = 0$) is not able to fully reproduce the atomistic results (ASD) for some all-optical switching processes in FiM samples. On the other hand, when such term is considered ($\vec{\tau}_{NE,i} \neq 0$), the extended LLB model (eLLB) naturally reproduces the atomistic temporal variation of the magnetization in small FiM samples under ultra-short laser pulses. This is shown in FIG. S4 for the same study as in Fig. 1(b) of the main text. Contrary to the ASD and the eLLB models, the conventional LLB ($\vec{\tau}_{NE,i} = 0$) does not predicts switching for $Q = 12 \times 10^{21}$ W/m³ pulses (Fig. S4(c)-(d)). Moreover, as shown in FIG. S4(a)-(b), the extended LLB model (eLLB), which includes $\vec{\tau}_{NE,i} \neq 0$, reproduces similar results even in the absence of thermal fluctuations ($\vec{\xi}_i^{\parallel} = \vec{\xi}_i^{\perp} = 0$).

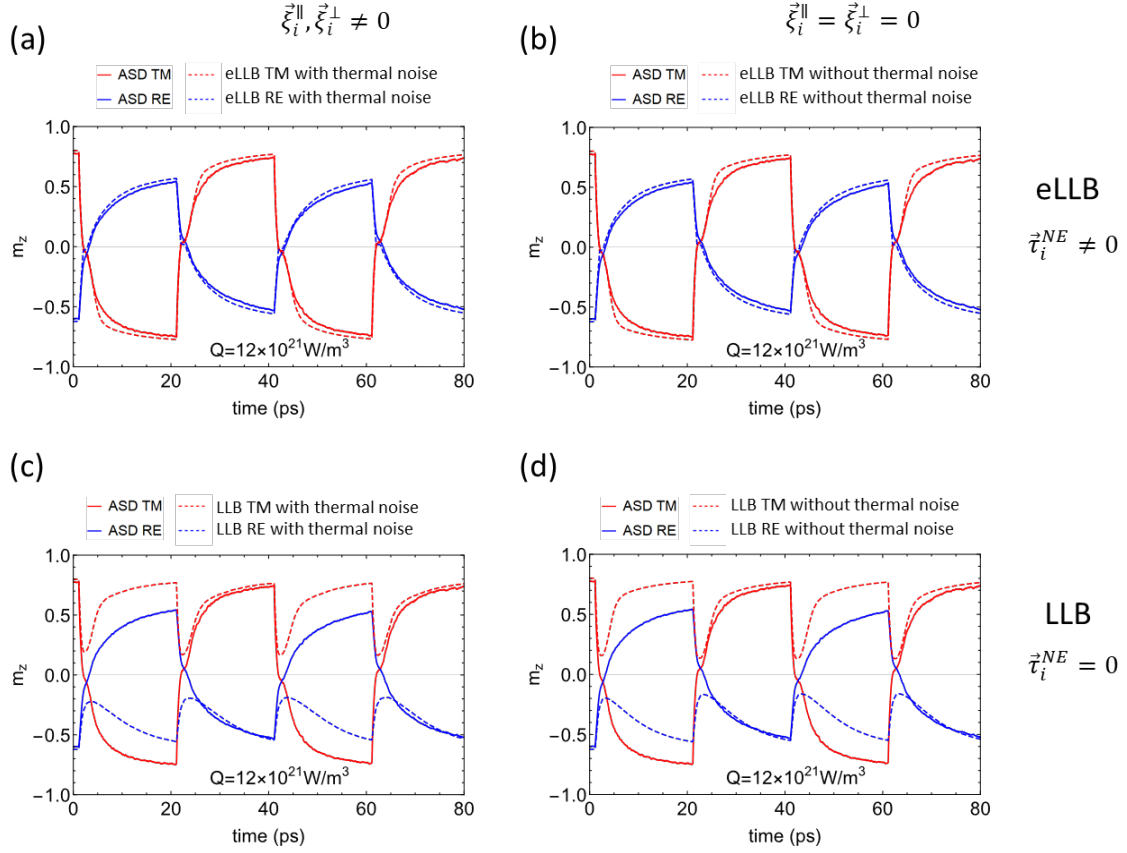

FIG. S4. Temporal evolution of the out-of-plane component ($m_z$ vs $t$) calculated by three different models: ASD, conventional LLB and extended eLLB models for the case studied in Fig. 1(b) of the main text. Solid lines in all graphs correspond to the ASD results. Red and blue curves correspond to the TM and RE respectively. In top graphs, dashed lines correspond to extended eLLB results ($\vec{\tau}_i^{NE} \neq 0$): (a) including thermal fluctuations $\vec{\xi}_i^{\parallel}, \vec{\xi}_i^{\perp} \neq 0$, and (b) without thermal fluctuations ($\vec{\xi}_i^{\parallel} = \vec{\xi}_i^{\perp} = 0$, deterministic case). In bottom graphs, dashed lines correspond to LLB results ($\vec{\tau}_i^{NE} = 0$): (a) including thermal fluctuations $\vec{\xi}_i^{\parallel}, \vec{\xi}_i^{\perp} \neq 0$, and (b) without thermal fluctuations ($\vec{\xi}_i^{\parallel} = \vec{\xi}_i^{\perp} = 0$, deterministic case).

**(c)** Except the contrary is indicated, all presented results with the extended eLLB model were computed with $\lambda_{ex} = 0.013$. FIG. S5 show the comparison of ASD results to eLLB results for three different values of $\lambda_{ex}$. A good quantitative agreement in the time traces of the out-of-plane components of each sublattice predicted by the ASD model is achieved time when $\lambda_{ex} = 0.013$ is considered (central graph in FIG. S5). For a laser pulse of $\tau_L = 50$ fs, and using $\lambda_{ex} = 0.013$, the threshold laser power density to achieve the switching is $Q = 7.9 \times 10^{21}$ W/m³ within the ASD, whereas the threshold is $Q = 9.9 \times 10^{21}$ W/m³ with the eLLB model. However, if $\lambda_{ex} = 0.024$ both models give the same threshold ($Q = 9.9 \times 10^{21}$ W/m³).

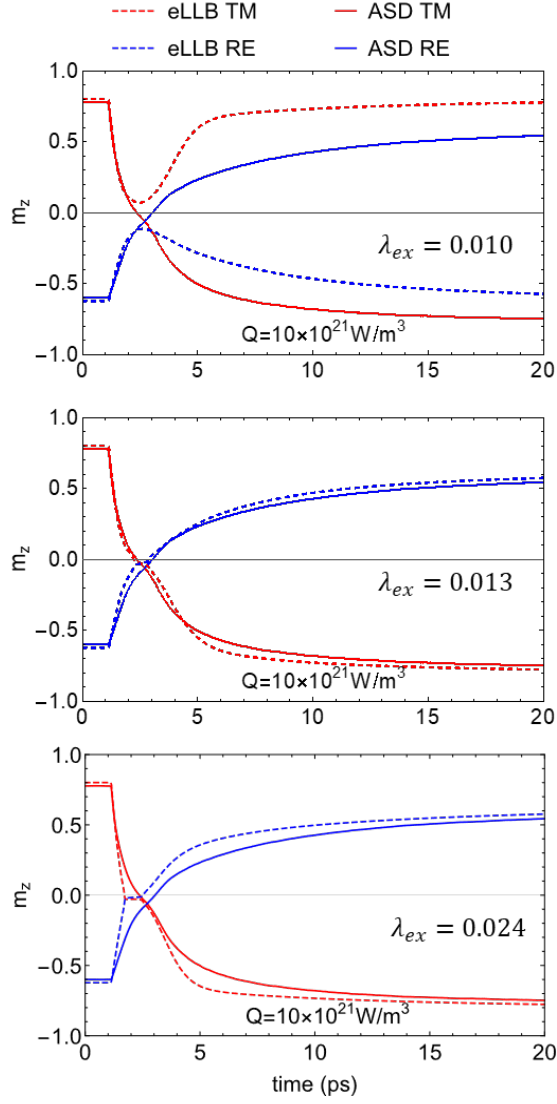

FIG. S5. Temporal evolution of the out-of-plane component ($m_z$ vs $t$) calculated with three different values of $\lambda_{ex}$ as indicated within each graph. Here $Q = 10 \times 10^{-21}$ W/m$^3$. The rest of inputs are the same as in FIG. S4.

**SN4.** *Phase diagram in terms of the fluence and the pulse duration*

FIG. 2(a) of the main text presents the phase diagram of the three possible final states as function of the pulse length ($\tau_L$) and the absorbed energy from the laser pulse ($Q$). Same results can be also presented in terms of the absorbed fluence, which is given by $F = Q\tau_L t_{FiM}$, where $t_{FiM}$ is the thickness of the ferrimagnetic alloy. This is shown in FIG. S6 for an extended range of laser pulse durations ($\tau_L$). These results are also in agreement with recent experimental studies [11]

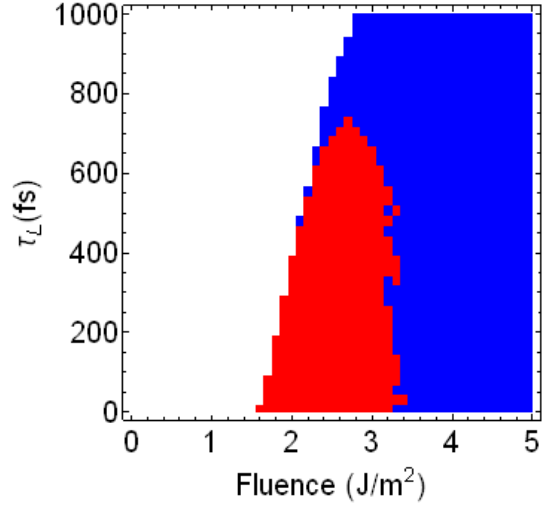

FIG. S6. Same results as FIG. 2(a) of the main text showing the three possible final states as a function of the fluence, $F = Q\tau_L t_{FiM}$ for an extended range of pulse durations ($\tau_L$). White, red and blue colors represent no-switching, deterministic switching (HI-AOS) and thermal demagnetization behaviors respectively.

**SN5.** *Helicity-Dependent AOS (HD-AOS) and Magnetic Circular Dichroism (MCD)*

Considering the Magnetic Circular Dichroism (MCD) scenario, the absorbed power by the FiM under circular polarized laser pulses, $P(r,t)$, also depends on the magnetic state of the system and the helicity of the pulse. Therefore, $P(r,t)$ in Eq. (2) of the TTM is replaced by $\psi(\sigma^\pm, m_N)P(r,t)$, where $\psi(\sigma^\pm, m_N)$ describes the different absorption power for *up* ($m_N > 0, \uparrow$) and *down* ($m_N < 0, \downarrow$) magnetization states as depending on the laser helicity ($\sigma^\pm$) of the laser pulse,

$$\psi(\sigma^\pm, m_N) = \left(1 + \frac{1}{2}k\sigma^\pm \text{sign}(m_N)\right) \quad (eS22)$$

where $m_N$ is the local net out of plane magnetization, ($m_N = M_s^{TM} m_z^{TM} + M_s^{RE} m_z^{RE}$), $\sigma^\pm = \pm 1$ is the helicity of the laser pulse ($\sigma^+ = +1$ for right-handed helicity and $\sigma^- = -1$ left-handed helicity), and sign($m_N$) is the sign of the initial net magnetization. Note that $m_z^{TM} = \pm 1$ corresponds to $m_z^{RE} = \mp 1$, but $M_s^{TM}$ and $M_s^{RE}$ are both positive. $k$ is a factor that determines the difference in the power absorption with respect to the linearly polarized case ($k = 0$). According to this criterium, a state with local net magnetization $m_N > 0$ ($\uparrow$: *up*) absorbs more energy for a right-handed laser $\sigma^+$ and less for left-handed helicity ($\sigma^-$) than in the linearly polarization case ($\sigma = 0$). The opposite happens starting from a state with local *down* net magnetization $m_N < 0$ ($\downarrow$: *down*). Considering $\psi(\sigma^\pm, m_N)P(r,t)$ in the TTM Eqs (2)-(3) of the main text, we obtain isothermal curves of $T = T_e = 1000$ K showing the transition from no-switching to helicity-dependent switching as a function of $Q$ and $\tau_L$ for different combinations of the laser helicity and the initial magnetic state under uniform illumination.

The results are shown in FIG. S7(a) and (b) for $k = 0.1$ and $k = 0.02$ respectively, which correspond to differences in the power absorption of 10% and 2% between *up* and *down* states. Solid black line corresponds to the case of linear polarization ($\sigma = 0$) already plotted in Fig. 2(a) of the main text. Cases with $(\sigma^+, \uparrow)$ and $(\sigma^-, \downarrow)$ are represented by the solid-blue curve, whereas solid-red curve corresponds to cases with $(\sigma^+, \downarrow)$ and $(\sigma^-, \uparrow)$. As it can be observed for $k = 0.1$ (FIG. S7(a)), the isothermal curves $T = T_e = 1000$ K for combinations $(\sigma^+, \uparrow)$ and $(\sigma^-, \downarrow)$ slightly reduce the values of $Q$ and $\tau_L$ needed to achieve switching with respect to the linear polarization case ($\sigma = 0$). On the contrary, the isothermal curve for combinations $(\sigma^+, \downarrow)$ and $(\sigma^-, \uparrow)$ slightly moves towards high values of $Q$ and $\tau_L$ with respect to linear polarization case ($\sigma = 0$). If $k = 0.02$ (FIG. S7(b)), the differences are even smaller, and the three curves almost overlap. This fact agrees with experimental observations, where the Helicity-Dependent AOS is only achieved in a narrow range of absorbed power ($Q$) for a fixed pulse length ($\tau_L$). Dashed lines in FIG. S7(a)-(b) correspond to the isothermal curves of $T = T_e = 1400$ K, and indicate the transition between switching and demagnetized multidomain behaviors.

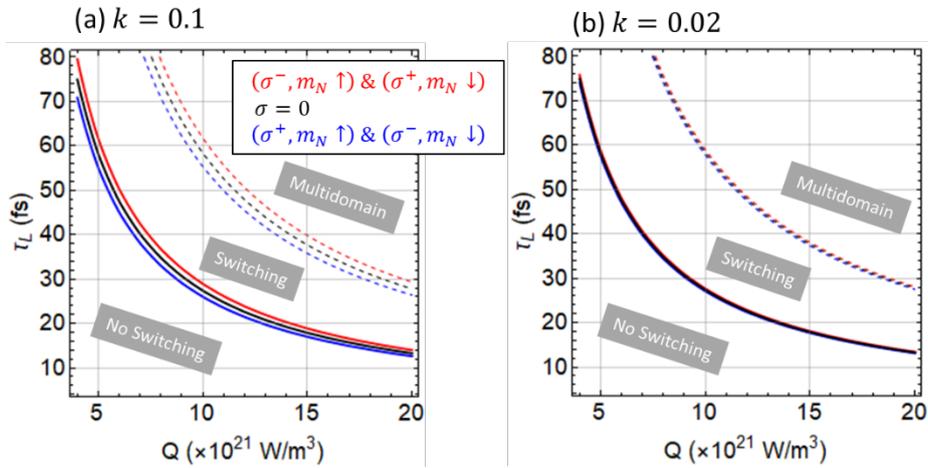

FIG. S7. Phase diagram in the MCD scenario for different values of the MCD coefficient $k$. (a) and (b) show the isothermal curves of $T = T_e = 1000$ K (solid lines) as a function of $Q$ and $\tau_L$ indicating the transition between no-switching to switching, as computed from the TTM Eq. (2)-(3) for $k = 0.1$, and $k = 0.02$ respectively. Dashed curves are isothermal curves indicating the transition between the switching to multidomain regimes ($T_e = 1400$ K). Black curve corresponds to the case of linear polarization ($\sigma = 0$) already plotted in Fig. 2(a) of the main text. Cases with $(\sigma^+, \uparrow)$ and $(\sigma^-, \downarrow)$ are represented by the blue curve, whereas red curve corresponds to cases with $(\sigma^+, \downarrow)$ and $(\sigma^-, \uparrow)$. Here $\uparrow$ and $\downarrow$ represent the initial net magnetic state ($m_N$), either *up* or *down* respectively.

**SN6.** *Helicity-Dependent AOS (HD-AOS) and Inverse Fadaray Effect (IFE)*

As mentioned in the main text, several works claim that the observations of the HD-AOS can be ascribed to the **Inverse Faraday Effect** (IFE) [12]. Within this formalism, the laser pulse generates an effective out-of-plane magneto-optical field whose direction depends on the laser pulse helicity as $\vec{B}_{MO}(\vec{r}, t) = \sigma^{\pm} B_{MO} \eta(r) \epsilon(t) \vec{u}_z$, where $\eta(r) = \exp[-4\ln(2)\, r^2/(2r_0)^2]$ is the spatial field profile, and $\epsilon(t)$ is its temporal profile. Note that there the spatial dependence of $\vec{B}_{MO}(\vec{r}, t)$ is the same as the one of the absorbed power density. According to the literature [12], the so-called magneto-optical field $\vec{B}_{MO}(\vec{r}, t)$ has some persistence with respect to the laser pulse, and therefore its temporal profile is different for $t < t_0$ and $t > t_0$: $\epsilon(t < t_0) = \exp[-4\ln(2)\,(t - t_0)^2/\tau_L^2]$, and $\epsilon(t \geq t_0) = \exp[-4\ln(2)(t-t_0)^2/(\tau_L + \tau_D)^2]$, where $\tau_D$ is the delay time of the $\vec{B}_{MO}(\vec{r}, t)$ with respect to the laser pulse. We have evaluated this scenario by including this field $\vec{B}_{MO}(\vec{r}, t)$ in the effective field of Eq. (4). The results for the same FiM alloy considered up to here ($Gd_x(FeCo)_{1-x}$, with $x$=0.25), (see Supplemental Note SN5), are shown in FIG. S8 by considering a maximum magneto-optical field of $B_{MO} = 20$ T with a delay time of $\tau_D = \tau_L$.

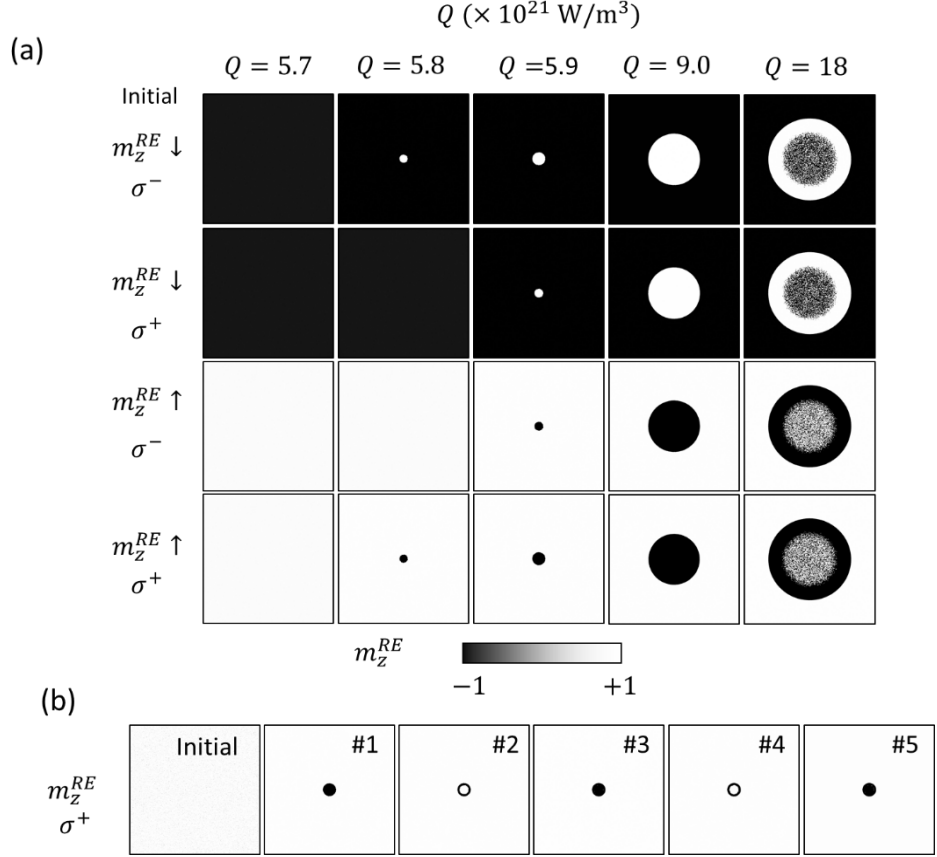

FIG. S8. Micromagnetic results of the HD-AOS computed within the IFE scenario. (a) Snapshots of the final RE magnetic state ($m_z^{RE}$) after a laser pulse of $\tau_L = 50$ fs for four different values of the absorbed power density ($Q$). Results are shown for four combinations of the initial state (↑,↓) and helicities ($\sigma^{\pm}$) as indicated at the left side. (b) RE magnetic state after every pulse with $\sigma^+$ for $Q = 5.9 \times 10^{21}$ W/m³, showing the appearance of a ring around the central part. The sample side is $\ell = 20$ μm and the laser spot diameter is $d_0 = \ell/2$. The IFE was evaluated by considering a maximum magneto-optical field of $B_{MO} = 20$ T with a delay time of $\tau_D = \tau_L$. The material parameters correspond to a FiM alloy $Gd_x(CoFe)_{1-x}$ with $x$=0.25.

**SN7.** *Inverse Faraday Effect: magneto-optical field or induced magnetic moment*

In previous section SN6 and in the main text, the Inverse Faraday Effect (IFE) was considered by adding an effective out-of-plane magneto-optical field ($\vec{B}_{MO}(\vec{r}, t)$) whose direction depends on the laser pulse helicity. Other authors [13,14] have alternatively pointed out that under circularly polarized laser pulses, the IFE can be taken into account in micromagnetic simulations by adding a helicity-dependent induced magnetic moment on each sublattice ($\Delta \vec{m}_i$). We have also explored this alternative by adding, instead of the $\vec{B}_{MO}(\vec{r},t)$, an additional term to the right-hand side of the eLLB Eq. (eS7). This term can be expressed as $\gamma'_0 \Delta \vec{m}_i$, where $\Delta \vec{m}_i$ represents the laser induced magnetization for sublattice $i$: RE, TM, and it is given by

$$\Delta \vec{m}_i(\vec{r},t) = (\sigma^{\pm}) \frac{K_{IFE,i} I}{c} \vec{u}_z \qquad (eS23)$$

where $K_{IFE,i}$ is the IFE constant for sublattice $i$: RE, TM, $I = P(r,t) t_{FiM}$ is the laser intensity with $P(r,t) = Q\eta(r)\xi(t)$ (see main text), and $c$ the speed of light. In FIG. S9 we compare both options to account for the IFE: either by a magneto-optical field ($\vec{B}_{MO}(\vec{r},t) \neq 0$ with $\Delta \vec{m}_i(\vec{r},t) = 0$, solid lines in FIG. S9) as described in SN6, or by an induced magnetic moment ($\vec{B}_{MO}(\vec{r},t) = 0$ with $\Delta \vec{m}_i(\vec{r},t) \neq 0$, dashed lines in FIG. S9). As it can be observed, both alternatives provide very similar results, so we conclude, as in [14], that both alternatives are essentially equivalent within the scope of our numerical study. Moreover, the adopted value of $\sigma^{\pm} K_{IFE,i} = \pm 0.0014$ T$^{-1}$ (see FIG. S9) is also in good quantitative agreement with typical values deduced in [14] from an *ab initio* formalism.

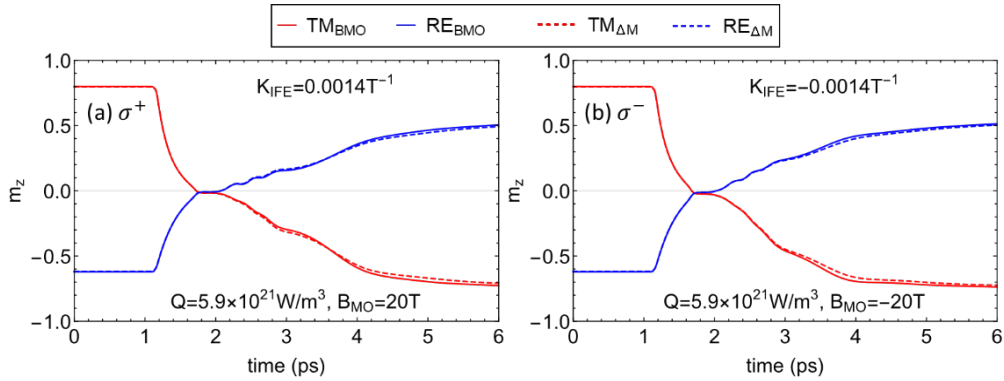

FIG. S9. Comparison of two micromagnetic implementations of the IFE for both laser helicities: (a) $\sigma^{+} = +1$ and (b) $\sigma^{-} = -1$. Solid lines correspond to results obtained assuming that the IFE generates a magneto-optical field ($\vec{B}_{MO}(\vec{r},t) \neq 0$), whereas dashed lines are for the induced magnetic moment alternative ($\Delta \vec{m}_i(\vec{r},t) \neq 0$). The values of the maximum magneto-optical field is $B_{MO} = 20$ T, whereas the IFE constant is $\sigma^{\pm} K_{IFE,i} = \pm 0.0014$ T$^{-1}$. The rest of inputs are the same as in Fig. S8.

**SN8.** *Material inputs for two different compositions*

In the last part of the manuscript we studied the probability of switching (see FIG. 6) for two FiMs (Gd$_x$(CoFe)$_{1-x}$:RE$_x$(TM)$_{1-x}$) with two different compositions: $x=0.25$ and $x=0.24$. The inputs used in these simulations are collected in the following Table S1.

|  | $x=0.25$: Gd$_x$(CoFe)$_{1-x}$ | $x=0.24$: Gd$_x$(CoFe)$_{1-x}$ |
|---|---|---|
| $M_{s,TM}$ / $M_{s,RE}$ (MA/m) | 0.412 / 0.546 | 0.412 / 0.52 |
| $A_{ex,TM}$ / $A_{ex,RE}$ (pJ/m) | 3.27 / 0.189 | 3.35 / 0.174 |
| $J_{TM-TM}$ / $J_{RE-RE}$ ($\times 10^{-21}$ J) | 2.59 / 1.35 | 2.59 / 1.35 |
| $J_{TM-RE}$ ($\times 10^{21}$ J) | -1.12 | |
| $K_{u,TM}$ / $K_{u,RE}$ (MJ/m$^3$) | 1.87 / 0.62 | 1.89 / 0.59 |
| $\mu_{TM}$ / $\mu_{RE}$ ($\mu_B$) | 1.92 / 7.63 | 1.92 / 7.63 |
| $\alpha_{TM}$ / $\alpha_{RE}$ ( ) | 0.02 / 0.02 | 0.02 / 0.02 |
| $\gamma_{TM}$ / $\gamma_{RE}$ ($10^{11} \times (T \cdot s)^{-1}$) | 1.847 / 1.759 | 1.847 / 1.759 |
| $g_{TM}$ / $g_{RE}$ ( ) | 2.1 / 2.0 | 2.1 / 2.0 |
| $a$ (nm) | 0.352 | |
| | | |
| $k_e$ (W/ (K · m)) | 91 | |
| $C_e$(300 K) ($\times 10^5$ J/ (K · m$^3$)) | 1.8 | |
| $C_l$ ($\times 10^6$ J/ (K · m$^3$)) | 3.8 | |
| $g_{el}$(300 K) ($\times 10^{17}$ W/m$^3$) | 7 | |

TABLE S1. Material inputs adopted to explore the switching probability for two different FiM alloys, with different compositions *x*. These inputs were used to get the results of FIG. 6 in the main text.

## SN9. *Helicity-Dependent All Optical Switching: MCD & IFE for different compositions and initial temperatures*

In FIG. 6 of the main text we explore the switching probability predicted by both the MCD and IFE mechanisms for two different compositions of the FiM and considering that the initial temperature of the thermal bath was room temperature. Similar results can be also obtained by fixing the composition of the FiM alloy and changing the temperature of the thermal bath with a cryostat. These results are shown in FIG. S10 for with *x*=0.25 (left column) and with *x*=0.24 (right column) compositions and three different temperatures of the thermal bath: $T = 260$ K, $T = 300$ K, and $T = 340$ K. For *x*=0.25, both the IFE and MCD predict similar behavior for $T = 260$ K (FIG. S10(b)) and $T = 300$ K (FIG. S10(c)): respect to the linear polarized laser pulse, the 100% switching probability occurs with smaller $Q$ for circular polarization when ($\sigma^-, m_z^{TM} \uparrow$) and ($\sigma^+, m_z^{TM} \downarrow$), and with larger $Q$ when ($\sigma^+, m_z^{TM} \uparrow$) and ($\sigma^-, m_z^{TM} \downarrow$). However, at $T = 340$ K (FIG. S10(d)), the IFE scenario results in similar behavior but the MCD results are the opposite. This can be easily understood as explained in the main, because the FiM alloy with $x = 0.25$ is RE-dominated for $T = 260$ K and $T = 300$ K, whereas becomes TM-dominated for $T = 340$ K. (Fig. S8(a)). Note that the magnetization compensation temperature for $x = 0.25$ is $T_M \approx 320$ K.

Similar results are also achieved for $x = 0.24$ (right column in FIG. S10), but now both IFE and MCD only give similar results for $T = 260$ K (FIG. S10(f)), whereas they predict opposite behavior for $T = 300$ K (FIG. S10(g)) and $T = 300$ K (FIG. S10(h)). Note that for $x = 0.24$, the FiM is only RE-dominated for temperature below the $T_M$, which now is $T_M \approx 280$ K.

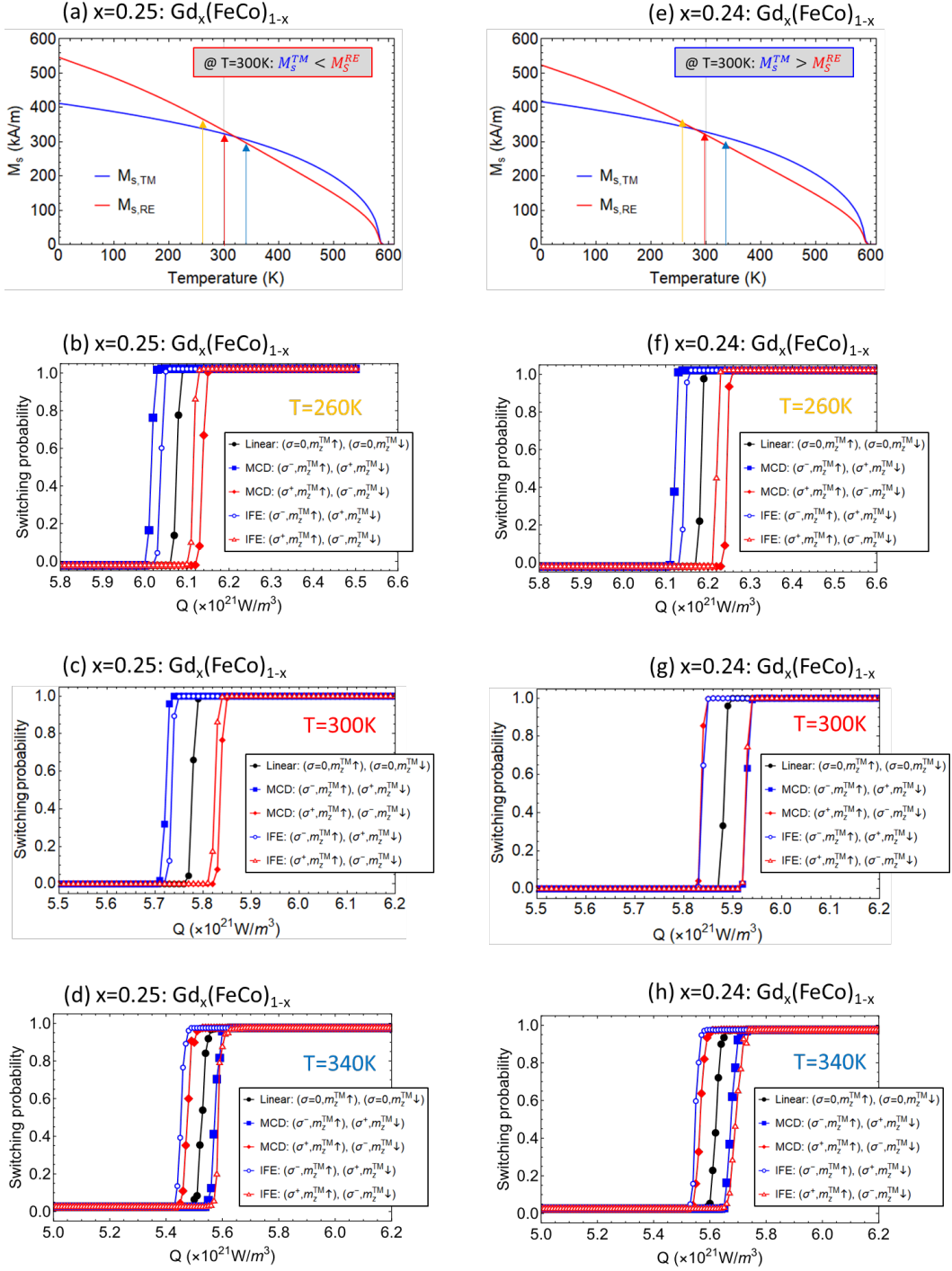

FIG. S10. Temperature dependence of the spontaneous magnetization of each sublattice (RE:Gd; TM:CoFe) of the FiM alloy ($Gd_x(CoFe)_{1-x}$) for two different compositions: (a) $x$=0.25 and (e) $x$=0.24. Probability of switching as a function of the absorbed power density ($Q$) for a laser pulse of $\tau_L = 50$ fs for different combinations of the initial state ($m_z^{TM}$: ($\uparrow, \downarrow$)) and the polarization (linear: $\sigma = 0$ (black dots), and circular $\sigma^\pm = \pm 1$) of the laser pulse as indicated in the legend and in the main text: (b), (c) and (d) corresponds for $T = 260$ K, $T = 300$ K, and $T = 340$ K for $x$=0.25, whereas (f), (g) and (h) to $x$=0.24. MCD results are shown by solid dots, whereas IFE results are presented by open symbols. Lines are guide to the eyes.